\documentclass[superscriptaddress,twocolumn,secnumarabic,floatfix,
nobibnotes,aps,prd,nofootinbib,10pt,longbibliography,noeprint]{revtex4-2}

\usepackage{comment}
\usepackage{graphicx}
\usepackage{bm}
\usepackage{amsmath}
\usepackage{amsfonts}
\usepackage{amssymb}
\usepackage{natbib}
\usepackage[dvipsnames]{xcolor}
\usepackage{verbatim}
\usepackage{soul}
\usepackage{physics}
\usepackage{microtype}
\usepackage[colorlinks=true,
            linkcolor=blue,
            urlcolor=blue,
            citecolor=blue,
            anchorcolor=blue]{hyperref}
\usepackage{orcidlink}
\usepackage[capitalize]{cleveref}
\crefname{section}{Sec.}{Secs.}
\crefname{figure}{Fig.}{Figs.}
\crefname{footnote}{footnote}{footnotes}
\usepackage[normalem]{ulem}
\usepackage{enumitem}
\usepackage{booktabs}
\usepackage{multirow}

\providecommand{\U}[1]{\protect\rule{.1in}{.1in}}

\newcommand{\be}{\begin{equation}}
\newcommand{\ee}{\end{equation}}
\newcommand{\bea}{\begin{eqnarray}}
\newcommand{\eea}{\end{eqnarray}}

\makeatletter
\def\Dated@name{}
\def\@bibstyle{apsrev4-2-prd}
\makeatother

\hypersetup{pdftitle={Exploring the growth-index (γ) tension with ΛsCDM},
pdfauthor={Luis A. Escamilla; Özgür Akarsu; Eleonora Di Valentino; Emre Özülker; J. Alberto Vazquez}}
\begin{document}

\date{}

\title{Exploring the growth-index ($\gamma$) tension with $\Lambda_{\rm s}$CDM}

\author{\mbox{Luis A. Escamilla}\,\orcidlink{0000-0003-4334-5140}}
\email[Contact author: ]{torresl@itu.edu.tr}
\affiliation{Department of Physics, Istanbul Technical University, Maslak 34469 Istanbul, Turkey}
\affiliation{School of Mathematical and Physical Sciences, University of Sheffield, Hounsfield Road, Sheffield S3 7RH, United Kingdom}

\author{\mbox{\"{O}zg\"{u}r Akarsu}\,\orcidlink{0000-0001-6917-6176}}
\email[Contact author: ]{akarsuo@itu.edu.tr}
\affiliation{Department of Physics, Istanbul Technical University, Maslak 34469 Istanbul, Turkey}

\author{\mbox{Eleonora Di Valentino}\,\orcidlink{0000-0001-8408-6961}}
\email[Contact author: ]{e.divalentino@sheffield.ac.uk}
\affiliation{School of Mathematical and Physical Sciences, University of Sheffield, Hounsfield Road, Sheffield S3 7RH, United Kingdom}

\author{\mbox{Emre \"{O}z\"{u}lker}\,\orcidlink{0000-0003-0817-4219}}
\email[Contact author: ]{e.ozulker@sheffield.ac.uk}
\affiliation{School of Mathematical and Physical Sciences, University of Sheffield, Hounsfield Road, Sheffield S3 7RH, United Kingdom}

\author{\mbox{J. Alberto Vazquez}\,\orcidlink{0000-0002-7401-0864}}
\email[Contact author: ]{javazquez@icf.unam.mx}
\affiliation{Instituto de Ciencias F\'isicas, Universidad Nacional Aut\'onoma de M\'exico, Cuernavaca, 
Morelos 62210, M\'exico}

\begin{abstract}
Recent observational analyses have revealed a significant tension in the growth index $\gamma$, which characterizes the growth rate of cosmic structures. When $\gamma$ is treated as a free parameter within lambda cold dark matter ($\Lambda$CDM), the PL18+$f\sigma_8$ combination yields $\gamma \approx 0.64$, in $\sim4\sigma$ tension with the theoretically expected value $\gamma \approx 0.55$ (assuming general relativity), suggesting a preference for suppressed late-time growth relative to the standard expectation. In this work, we investigate whether the $\Lambda_{\rm s}$CDM framework---featuring a rapid sign-switching cosmological constant (mirror anti--de Sitter-to-de Sitter (AdS-to-dS) transition) in the late universe at redshift $z_\dagger \sim 2$---can alleviate the $\gamma$ tension together with the $H_0$ and $S_8$ tensions. 
We find that $\gamma\Lambda_{\rm s}$CDM is able to reduce the $\gamma$ and $H_0$ tensions relative to $\gamma\Lambda$CDM while maintaining low $S_8$ values. For PL18+$f\sigma_8$, the combination underlying the $\gamma$ tension, $\gamma\Lambda_{\rm s}$CDM yields $z_\dagger>1.58$ (95\% CL) and reduces the $\gamma$ and $H_0$ discrepancies from $3.7\sigma$ and $4.2\sigma$ in $\gamma\Lambda$CDM to $2.1\sigma$ and $1.1\sigma$, respectively, while leaving $\chi^2_{\rm min}$ essentially unchanged. The largest improvement among the free-$z_\dagger$ analyses is obtained for PL18+$f\sigma_8$+PP\&SH0ES, for which the marginalized posterior constraint is $z_\dagger=1.82^{+0.32}_{-0.28}$ (95\% CL); the $\gamma$ and $H_0$ tensions are reduced to $1.8\sigma$ and $0.6\sigma$, respectively, and the model also provides an improved fit to the data. 
The efficacy of the mechanism is, however, dataset dependent: the standard 3D BAO measurements used in the present analysis---from SDSS-IV/eBOSS and the first year of DESI observations---tend to push the transition to higher redshifts, $z_\dagger\gtrsim2.3$--$2.4$, and weaken the simultaneous alleviation, particularly of the $H_0$ tension. We also examine the benchmark case with fixed $z_\dagger=1.7$, previously identified in studies of the standard ($\gamma\simeq0.55$) $\Lambda_{\rm s}$CDM framework as a redshift of interest for alleviating multiple major cosmological tensions---including those in $H_0$, $M_{\rm B}$, and $S_8$. For the restricted combinations in which it is considered, this benchmark yields low residual tensions (e.g., a $1.6\sigma$ $\gamma$ tension and an effectively vanishing $H_0$ tension for PL18+$f\sigma_8$), albeit with somewhat worse fits than $\gamma\Lambda$CDM. It is, however, excluded at more than $2\sigma$ by the free-$z_\dagger$ analyses once either the uncalibrated PantheonPlus SN~Ia sample or the 3D BAO data are included. Our results indicate that $\Lambda_{\rm s}$CDM can contribute to the joint alleviation of multiple cosmological tensions, although its impact is dataset dependent.
\end{abstract}

\maketitle

\section{Introduction}

It is well established that, despite its status as the prevailing cosmological paradigm, the standard lambda cold dark matter ($\Lambda$CDM) model faces several important observational challenges and inconsistencies, which appear to be intensifying as the cosmological data become more precise. Some of these issues relate to fundamental gaps in our understanding of the dark sector, comprising dark matter (DM) and dark energy (DE). Others are statistical in nature, but equally significant, most notably the well-documented $H_0$ (Hubble constant) tension~\cite{Verde:2019ivm,DiValentino:2020zio,DiValentino:2021izs,Perivolaropoulos:2021jda,Schoneberg:2021qvd,Shah:2021onj,Abdalla:2022yfr,DiValentino:2022fjm,Kamionkowski:2022pkx,Hu:2023jqc,Giare:2023xoc,Verde:2023lmm,DiValentino:2024yew,Perivolaropoulos:2024yxv,Akarsu:2024qiq,CosmoVerseNetwork:2025alb}, and the $S_8$ (weighted amplitude of matter fluctuations) tension~\cite{DES:2021wwk,KiDS:2020suj,Heymans:2020gsg,Kilo-DegreeSurvey:2023gfr,Dalal:2023olq,Chen:2024vvk,Kim:2024dmg,DES:2024oud,Harnois-Deraps:2024ucb,Qu:2024sfu,DiValentino:2018gcu,Troster:2019ean,DiValentino:2020vvd,Adil:2023jtu,Akarsu:2024hsu,Akarsu:2024qiq,Sailer:2024jrx,CosmoVerseNetwork:2025alb}.  
In addition to these established discrepancies, recent analyses have sharpened a tension associated with the growth rate of cosmic structures (often discussed for over a decade in the context of ``low-growth'' indications from large-scale structure~\cite{Macaulay:2013swa,Nesseris:2017vor,Gomez-Valent:2018nib,Benisty:2020kdt} and, more recently, recast as a growth-index ($\gamma$) tension~\cite{Nguyen:2023fip,Specogna:2023nkq}) relative to expectations within the standard $\Lambda$CDM framework.

The linear growth rate of matter density perturbations, defined as $f(a) \equiv \frac{d \ln \delta(a)}{d \ln a}$, where $\delta(a)$ is the matter density contrast and $a$ is the scale factor, characterizes the dynamical evolution of cosmic structures under gravitational instability. Within the framework of general relativity (GR), this quantity can be reliably approximated by the power-law ansatz $f(a) \approx \Omega_{\rm m}(a)^\gamma$~\cite{Wang:1998gt,Linder:2005in}, where $\Omega_{\rm m}(a)$ is the matter density parameter, and $\gamma$ is the growth index, a parameter that encapsulates the influence of cosmic expansion on structure formation. 
In standard $\Lambda$CDM cosmology, assuming a spatially flat universe and GR as the underlying theory of gravity, the growth index is predicted to be $\gamma \approx 0.55$ with high accuracy. Precise observational constraints on $f(a)$, particularly at low redshifts, provide a powerful probe of the nature of dark energy, the properties of dark matter, and potential deviations from GR on cosmological scales~\cite{Huterer:2022dds}. Any statistically significant departure from $\gamma \approx 0.55$ may signal the presence of new physics, such as modifications to gravity or interactions beyond the Standard Model (SM) of particle physics.

In Ref.~\cite{Nguyen:2023fip}, it was demonstrated that when the growth index $\gamma$ is treated as a free parameter within $\Lambda$CDM and inferred from large-scale structure and Planck cosmic microwave background (CMB) data, a tension exceeding $4\sigma$ arises relative to the theoretically expected value of $\gamma \approx 0.55$. This discrepancy might be linked to the well-known $A_{\rm lens}$ problem~\cite{Nguyen:2023fip, Specogna:2023nkq, Planck:2018vyg, Calabrese:2008rt, DiValentino:2020hov, Specogna:2024euz}. Given the worrisome longstanding tensions and newly emerging ones, such as the $\gamma$ tension,\footnote{It is important to note that many of the tensions are not completely independent, e.g., the $H_0$ and $M_{\rm B}$ tensions noted below. Similarly, the $\gamma$ tension is closely related to the $S_8$ tension.} within $\Lambda$CDM, it is natural to explore alternative frameworks that may provide a viable resolution. In this context, the $\Lambda_{\rm s}$CDM model emerged as a promising candidate.

The so-called $\Lambda_{\rm s}$CDM framework~\cite{Akarsu:2019hmw,Akarsu:2021fol,Akarsu:2022typ,Akarsu:2023mfb} stands out as a promising and economical extension of $\Lambda$CDM, capable of addressing major cosmological tensions---including those in $H_0$, $M_{\rm B}$ (type Ia supernovae absolute magnitude), and $S_8$---as well as the baryon acoustic oscillations (BAO) Ly-$\alpha$ anomaly, 
while simultaneously yielding an age of the Universe consistent with estimates from the oldest globular clusters and, when treated as free parameters, a total neutrino mass and an effective number of species in agreement with the SM of particle physics~\cite{Akarsu:2019hmw,Akarsu:2021fol,Akarsu:2022typ,Akarsu:2023mfb,Akarsu:2024qsi,Akarsu:2024eoo,Yadav:2024duq}.
 Due to the latter feature studied in~\cite{Yadav:2024duq}, we can speculate that the $\Lambda_{\rm s}$CDM framework may circumvent the recently proposed anomaly wherein cosmological data (e.g., DESI BAO), analyzed within the standard $\Lambda$CDM framework, appear to favor $m_\nu < 0$~\cite{Craig:2024tky,Wang:2024hen,Green:2024xbb,Herold:2024enb,Ge:2024kac,Jiang:2024viw}, as happens for other models of DE~\cite{Elbers:2024sha}. Motivated by this success of $\Lambda_{\rm s}$CDM, we explore $\Lambda_{\rm s}$CDM's potential to resolve the recently identified $\gamma$ tension~\cite{Nguyen:2023fip} by examining whether it naturally explains the observed suppression of cosmic structure growth while still remaining consistent with the theoretically expected value, $\gamma \approx 0.55$ (assuming GR). 
We first briefly review the $\Lambda_{\rm s}$CDM framework in Sec.~\ref{sec:mechanism} and provide insights into why it has the potential to address the $\gamma$ tension, along with the $H_0$ and $S_8$ tensions. Here, we recall only its defining feature: the cosmological constant $\Lambda$ is replaced by $\Lambda_{\rm s}$, which switches sign from a negative, anti--de Sitter (AdS)-like value at high redshifts to a positive, de Sitter (dS)-like value via a rapid \textit{mirror} (magnitude-preserving) AdS-to-dS transition at the transition redshift $z_\dagger\sim2$. The role of this transition redshift can already be understood from CMB-distance consistency. Since the model leaves the prerecombination Universe essentially unchanged, the sound horizon $r_*$ remains close to its Planck-$\Lambda$CDM value, while the precisely measured acoustic scale $\theta_*=r_*/D_M(z_*)$ fixes the comoving distance to last scattering and the CMB peak structure constrains the physical matter density $\Omega_{\rm m0}h^2$. 
Holding these CMB quantities close to their Planck-$\Lambda$CDM values and varying $z_\dagger$ therefore defines a CMB-compatible branch in the $H_0$--$z_\dagger$ plane. Along this branch, lowering $z_\dagger$ prolongs the AdS-like phase and thus suppresses $H(z)$ over a larger portion of the distance integral to last scattering; preserving $D_M(z_*)$ then requires a compensating enhancement of the late-time expansion rate. This leads to a strong anticorrelation between $z_\dagger$ and $H_0$, with transition redshifts around $z_\dagger\simeq1.5$--$1.7$ naturally yielding values of $H_0$ in the SH0ES range~\cite{Akarsu:2019hmw,Akarsu:2021fol}. This CMB-geometric argument, which we revisit in more detail in Sec.~\ref{sec:mechanism}, is what renders $z_\dagger\simeq1.7$ a phenomenologically motivated benchmark of the model: it is singled out by CMB consistency combined with the requirement of a SH0ES-level $H_0$.

Since CMB data alone leave a broad degeneracy in $z_\dagger$, late-time data determine which part of this CMB-compatible branch is selected. The standard three-dimensional BAO measurements used in the present analysis---those from SDSS-IV/eBOSS and the first year of DESI observations---tend to push the preferred transition to higher redshifts, $z_\dagger\gtrsim2.3$--$2.4$, thereby disfavoring the high-$H_0$, low-$\Omega_{\rm m0}$ portion of the branch. By contrast, transverse BAO (BAOtr, or 2D BAO) data, which are less dependent on the fiducial-cosmology distance conversion, have been shown in previous $\Lambda_{\rm s}$CDM analyses to select the low-transition-redshift regime, $z_\dagger<2$ (close to the benchmark $z_\dagger\simeq1.7$), and to preserve the associated alleviation of the $H_0$, $M_{\rm B}$, and $S_8$ tensions~\cite{Akarsu:2023mfb,Akarsu:2024eoo,Yadav:2024duq}. 
A preference for transition redshifts below the range selected by the standard 3D BAO data is also found in related and extended scenarios. In the $w$XCDM model---which likewise features an abrupt sign switch in the DE density, but allows the pre- and posttransition DE components to have independent EoS parameters $w_X$ and $w_Y$---transverse/2D and standard 3D BAO data again select different transition regimes, with BAOtr-inclusive fits yielding $z_t\simeq1.5$~\cite{Gomez-Valent:2024tdb,Gomez-Valent:2024ejh}. 
The string-inspired $\Lambda_{\rm s}$CDM$^+$ realization---which augments abrupt $\Lambda_{\rm s}$CDM with a theoretically predicted excess $\Delta N_{\rm eff}\simeq0.25$ in the effective number of relativistic species---yields a slightly higher $z_\dagger\simeq2.1$ in a BAOtr-inclusive analysis~\cite{Anchordoqui:2023woo,Anchordoqui:2024gfa,Anchordoqui:2024dqc,Soriano:2025gxd}. In the present work, whose aim is to parallel the growth-index analysis of Ref.~\cite{Nguyen:2023fip}, we use the standard 3D BAO datasets described below and recall the BAOtr/2D BAO results only as context for interpreting the fixed-$z_\dagger=1.7$ benchmark.

To this end, we extend both the abrupt $\Lambda_{\rm s}$CDM and $\Lambda$CDM models by treating $\gamma$ as a free parameter, constraining them using Planck CMB and $f\sigma_8$ data, and subsequently incorporating additional cosmological datasets such as type Ia supernovae (SN~Ia) and BAO data.\footnote{Throughout this work, BAO refers to the standard three-dimensional (3D) BAO measurements from the SDSS-IV/eBOSS and first-year DESI likelihoods described in Sec.~\ref{sec:MandD}; a more detailed discussion of the distinction between 3D and transverse (2D) BAO data is given in~footnote~\ref{footnote_2dbao}.}
The parameter $\gamma$ is introduced here as an effective, phenomenological diagnostic of the linear growth sector (in the same sense as in Ref.~\cite{Nguyen:2023fip}), not as a fundamental parameter. This allows us to assess the capability of $\Lambda_{\rm s}$CDM in addressing the $\gamma$ tension while simultaneously evaluating the status of the $H_0$ and $S_8$ tensions. Our analysis yields promising results and reveals a clear correlation in the alleviation of the $\gamma$ and $H_0$ tensions, reinforcing the possibility of a unified resolution to the major cosmological tensions within the $\Lambda_{\rm s}$CDM framework.

\section{Insights into $\Lambda_{\rm s}$CDM and Jointly Addressing the $\gamma$, $H_0$, and $S_8$ Tensions}
\label{sec:mechanism}

The $\Lambda_{\rm s}$CDM framework, originally conjectured phenomenologically in Ref.~\cite{Akarsu:2019hmw} based on findings from the graduated dark energy (gDE) model, posits that around redshift $z_\dagger \sim 2$, the Universe underwent a period of rapid \textit{mirror} AdS-to-dS (anti--de Sitter-to-de Sitter) transition in vacuum energy---namely, a rapid sign-switching cosmological constant (CC), $\Lambda_{\rm s}$, from negative to positive while preserving its overall magnitude (``mirror'' reflects this invariance)---or a similar phenomenon, all while leaving other standard cosmological components, such as baryons, CDM, prerecombination physics, and the inflation paradigm, unaltered. Such transitions can typically be described using sigmoidlike functions; for example, $\Lambda_{\rm s}(z) = \Lambda_{\rm s0} \tanh[\nu(z_{\dagger}-z)]/\tanh[\nu z_\dagger]$, where $\nu > 1$ controls the sharpness of the transition, $\Lambda_{\rm s0} > 0$ is the present-day value, and $z_{\dagger}$ denotes the transition redshift. 
For a sufficiently rapid transition, e.g., $\nu \gtrsim 10$ at $z_{\dagger} \sim 2$, this function approximates a smooth steplike behavior with $\Lambda_{\rm s} \approx \Lambda_{\rm s0}$ for $z \lesssim 2$ and $\Lambda_{\rm s} \approx -\Lambda_{\rm s0}$ for $z \gtrsim 2$, effectively characterized by a single parameter, $z_\dagger$~\cite{Akarsu:2024eoo}. In the limiting case $\nu \to \infty$, the transition becomes instantaneous:
\begin{equation}
\label{eqn:abruptlscdm}
\Lambda_{\rm s}(z) \to \Lambda_{\rm s0}\,{\rm sgn}(z_{\dagger}-z),
\end{equation}
which defines the \textit{abrupt} $\Lambda_{\rm s}$CDM model~\cite{Akarsu:2021fol,Akarsu:2022typ,Akarsu:2023mfb}---an idealized representation of a rapid mirror AdS-to-dS transition extending $\Lambda$CDM by a single additional parameter, $z_{\dagger}$. 
This is the simplest phenomenological realization of the $\Lambda_{\rm s}$CDM framework, which has been extensively studied~\cite{Akarsu:2021fol,Akarsu:2022typ,Akarsu:2023mfb,Yadav:2024duq,Akarsu:2024eoo} under the assumption that gravity is governed by GR. Its \textit{abrupt} transition at $z = z_{\dagger}$ introduces a discontinuity, resulting in a type II (sudden) singularity~\cite{Barrow:2004xh} at $z = z_\dagger$. Nonetheless, Ref.~\cite{Paraskevas:2024ytz} has demonstrated that this singularity itself has a negligible impact on bound cosmic structures and their formation and evolution, and it disappears when the transition is smoothed out. From a mathematical/physical standpoint, this framework remains identical to $\Lambda$CDM at redshifts $z < z_{\dagger}$, where it features a dS-like CC after the transition but introduces a minimal modification by incorporating an AdS-like CC for all redshifts prior to the transition, $z > z_{\dagger}$. 
However, from a phenomenological perspective---i.e., considering the expansion dynamics and observational signatures---the impact of this modification is effectively limited to redshifts $z \lesssim z_{\dagger} \sim 2$. Specifically, for $z < z_{\dagger}$, $\Lambda_{\rm s}$CDM closely replicates the expansion history $H(z)$ of $\Lambda$CDM, albeit with systematically higher values; it introduces a noticeable deformation of $H(z)$ around the transition redshift $z \sim z_{\dagger}$ and becomes virtually indistinguishable from $\Lambda$CDM at higher redshifts ($z \gtrsim 3$).
Consequently, from a phenomenological standpoint, $\Lambda_{\rm s}$CDM functions as a postrecombination, late-time modification of $\Lambda$CDM. For notable explicit physical realizations of this scenario---both in its abrupt and smooth formulations---see, for example, Refs.~\cite{Alexandre:2023nmh,Anchordoqui:2023woo,Anchordoqui:2024gfa,Anchordoqui:2024dqc,Akarsu:2024qsi,Akarsu:2024eoo,Akarsu:2024nas,Souza:2024qwd,Akarsu:2025gwi}, which explore its implementation within various well-founded theoretical frameworks.\footnote{For further theoretical and observational studies---including model-agnostic reconstructions---investigating DE (whether as a physical source or an effective source arising from, e.g., modified gravity) that features negative energy densities at high redshifts ($z \gtrsim 1.5\text{--}2$), often consistent with a negative (AdS-like) CC and aimed at addressing cosmological tensions, we refer the reader, without claiming to be exhaustive, to Refs.~\citep{Sahni:2002dx, Vazquez:2012ag, BOSS:2014hwf, Sahni:2014ooa, BOSS:2014hhw, DiValentino:2017rcr, Mortsell:2018mfj, Poulin:2018zxs, Capozziello:2018jya, Wang:2018fng, Banihashemi:2018oxo, Dutta:2018vmq, Banihashemi:2018has, Akarsu:2019ygx, Li:2019yem, Visinelli:2019qqu, Ye:2020btb, Perez:2020cwa, Akarsu:2020yqa, Ruchika:2020avj, DiValentino:2020naf, Calderon:2020hoc, Ye:2020oix, DeFelice:2020cpt, Paliathanasis:2020sfe, Bonilla:2020wbn, Acquaviva:2021jov, Bag:2021cqm, Bernardo:2021cxi, Escamilla:2021uoj, Sen:2021wld, Ozulker:2022slu, DiGennaro:2022ykp, Akarsu:2022lhx, Moshafi:2022mva, Bernardo:2022pyz, vandeVenn:2022gvl, Ong:2022wrs, Tiwari:2023jle, Malekjani:2023ple, Vazquez:2023kyx, Escamilla:2023shf, Adil:2023exv, Alexandre:2023nmh, Adil:2023ara, Paraskevas:2023itu, Gomez-Valent:2023uof, Wen:2023wes, Medel-Esquivel:2023nov, DeFelice:2023bwq, Anchordoqui:2023woo, Menci:2024rbq, Anchordoqui:2024gfa, Gomez-Valent:2024tdb, DESI:2024aqx, Bousis:2024rnb, Wang:2024hwd, Colgain:2024ksa, Tyagi:2024cqp, Toda:2024ncp, Sabogal:2024qxs, Dwivedi:2024okk, Escamilla:2024ahl, Anchordoqui:2024dqc, Akarsu:2024nas, Gomez-Valent:2024ejh, Manoharan:2024thb, Souza:2024qwd, Mukherjee:2025myk, Giare:2025pzu, Keeley:2025stf, Akarsu:2025gwi, Soriano:2025gxd,Efstratiou:2025xou}. 
Phantom DE models, traditionally assumed to yield only positive energy densities, are well known for alleviating the $H_0$ tension. Among these, the so-called \textit{phantom crossing model}~\cite{DiValentino:2020naf} (DMS20~\cite{Adil:2023exv}) stands out; a recent analysis~\cite{Adil:2023exv} reaffirming its success also revealed that its ability to assume negative densities for $z \gtrsim 2$---mimicking an AdS-like CC at higher redshifts---plays a central role in its effectiveness. Interacting DE (IDE) models~\cite{Kumar:2017dnp,DiValentino:2017iww,Yang:2018uae,Kumar:2019wfs,Pan:2019gop,DiValentino:2019ffd,DiValentino:2019jae,Lucca:2020zjb,Gomez-Valent:2020mqn,Kumar:2021eev,Nunes:2022bhn,Bernui:2023byc,Giare:2024smz,Giare:2024gpk,Giare:2024oil,Giare:2025pzu,Sabogal:2025mkp} provide an alternative approach to alleviating the $H_0$ tension, yet model-independent reconstructions of interaction kernels~\cite{Escamilla:2023shf} suggest their energy densities tend to become negative at $z \gtrsim 2$. 
Recent DESI BAO data---analyzed using the CPL parametrization---provided more than $3\sigma$ evidence for dynamical DE~\cite{DESI:2024mwx}, while nonparametric reconstructions of the same data further indicate that this dynamical behavior may involve vanishing or negative DE densities for $z \gtrsim 1.5\text{--}2$~\cite{DESI:2024aqx,Escamilla:2024ahl}, a trend similarly observed in pre-DESI BAO data from SDSS~\cite{Sabogal:2024qxs,Escamilla:2024ahl}.}

In this scenario, since the prerecombination Universe remains unaltered, as in $\Lambda$CDM, the comoving sound horizon at last scattering, $r_*$, remains effectively unchanged. The Planck CMB spectra provide precise measurements of the angular scale of the sound horizon, $\theta_* = r_*/D_M(z_*)$, and the present-day physical matter density, $\Omega_{\rm m0} h^2$ (here, $h=H_0/100\,{\rm km\,s}^{-1}\,{\rm Mpc}^{-1}$ is the dimensionless reduced Hubble constant), inferred from the peak structure and damping tail. 
Consequently, both the comoving angular diameter distance to last scattering, $D_M(z_*)$, and $\Omega_{\rm m0} h^2$ must remain consistent with their Planck-$\Lambda$CDM inferred values. 
Any suppression of $H(z)$ when $z > z_\dagger$, caused by the AdS-like CC, must therefore be compensated by an enhancement of $H(z)$ when $z < z_\dagger$ during the dS-like CC regime to preserve consistency with Planck-$\Lambda$CDM inferred $D_M(z_*)$. 
This mechanism naturally increases $H_0$ while lowering $\Omega_{\rm m0}$ relative to Planck-$\Lambda$CDM. A later transition (i.e., a smaller $z_\dagger$) prolongs the suppression of $H(z)$ and deepens it at later times, amplifying both effects, provided that the transition occurs before the AdS-like CC dominates and halts expansion. 
Additionally, the suppression of $H(z)$ for $z > z_{\dagger}$ reduces cosmic friction, thus enhancing structure formation in this regime, which could potentially lead to an increase in $\sigma_8$ (amplitude of mass fluctuations on scales of $8\,h^{-1}\,\rm Mpc$). 
However, the reduced present-day matter density parameter can still lower $S_8 = \sigma_8 \sqrt{\Omega_{\rm m0}/0.3}$ and thereby alleviate the $S_8$ tension, particularly when the transition occurs sufficiently late (roughly $z_\dagger\lesssim 2$). Similarly, since the linear growth rate of matter perturbations $f$ is related to the growth index $\gamma$ through $f(a) \approx \Omega_{\rm m}(a)^\gamma$, with $\gamma \approx 0.55$ in GR, a late-time suppression in $f$ can be achieved through reduced values of $\Omega_{\rm m}(a)$ relative to $\Lambda$CDM at $z < z_{\dagger}$ while still maintaining $\gamma \sim 0.55$---thus making the same mechanism responsible for alleviating the $S_8$ tension a compelling candidate to address the $\gamma$ tension as well.

A theoretical study of the abrupt $\Lambda_{\rm s}$CDM model in Ref.~\cite{Akarsu:2025ijk}, within GR, reveals a two-phase evolution of linear matter density perturbations. Before the transition, during the AdS-like CC regime, the negative cosmological constant reduces cosmic friction, enhancing the growth rate of matter perturbations relative to $\Lambda$CDM, with the effect peaking just before the transition. After the transition, the model closely follows $\Lambda$CDM but with a larger dS-like CC, leading to higher expansion rates $H(z)$ and stronger cosmic friction, which more effectively suppresses structure growth at late times. Specifically, before the transition, the growth index $\gamma$ remains below the matter-dominated limiting value in $\Lambda$CDM ($\gamma \to 6/11$)~\cite{Wang:1998gt} and thus lower than in $\Lambda$CDM. 
As the transition occurs, $\gamma$ rises sharply and then evolves gradually, closely tracking the $\Lambda$CDM prediction while staying slightly elevated posttransition. Nevertheless, $\gamma$ remains near $0.55$ throughout, consistent with standard $\Lambda$CDM expectations; see Fig.~11 in Ref.~\cite{Akarsu:2025ijk}. Using $\gamma= 0.55$ and the Planck-$\Lambda$CDM prediction of $\Omega_{\rm m0} \simeq 0.32$, the corresponding present-day growth rate is $f_0 \simeq 0.53$. 
However, Nguyen et al.~\cite{Nguyen:2023fip}, extending $\Lambda$CDM by treating $\gamma$ as a free parameter constrained by observational data (PL18+$f\sigma_8$), report $\gamma \simeq 0.64$, leading to a suppressed growth rate of $f_0 \simeq 0.47$ (cf.~\cref{tab:results}), indicating a deviation from $\Lambda$CDM and a suppression of structure growth at low redshifts. Observational analyses show that $\Lambda_{\rm s}$CDM generally predicts lower $\Omega_{\rm m0}$ than $\Lambda$CDM~\cite{Akarsu:2021fol,Akarsu:2022typ,Akarsu:2023mfb,Akarsu:2024eoo}; specifically, Planck-$\Lambda_{\rm s}$CDM yields $\Omega_{\rm m0} \simeq 0.28$, corresponding to $ f_0 \simeq 0.49 $, closely matching Ref.~\cite{Nguyen:2023fip}. Moreover, it was shown that for $z_\dagger\simeq1.7$, $\Lambda_{\rm s}$CDM resolves both the $H_0$ and $S_8$ tensions, bringing them below $\sim1\sigma$~\cite{Akarsu:2023mfb} by predicting $\Omega_{\rm m0} \simeq 0.27$, which in turn gives $ f_0 \simeq 0.48 $, in precise agreement with the inferred $\gamma$ in~\cite{Nguyen:2023fip}.

This trend parallels the way $\Lambda_{\rm s}$CDM alleviates the $S_8$ tension, primarily through subtle modifications in the CMB constraints, effectively reducing the discrepancy from $3.1\sigma$ in $\Lambda$CDM to approximately $1.7\sigma$. Notably, since $\Lambda_{\rm s}$CDM is identical to $\Lambda$CDM after the transition (for $z \lesssim z_\dagger \sim 2$), its KiDS-1000-alone constraint closely matches that of $\Lambda$CDM, yielding $S_8 = 0.746^{+0.026}_{-0.021}$ compared to $S_8 = 0.749^{+0.027}_{-0.020}$ in $\Lambda$CDM. However, in the Planck-alone analysis, $\Lambda_{\rm s}$CDM predicts only slightly larger $\sigma_8$ than $\Lambda$CDM. Consequently, by significantly reducing $\Omega_{\rm m0}$---and given the positive correlation between $S_8$ and $\Omega_{\rm m0}$ (which itself depends on the transition redshift $z_\dagger$)---the model naturally shifts $S_8$ to lower values, improving agreement with its KiDS-1000-alone constraints. 
Specifically, the Planck-alone analysis yields $S_8 = 0.801^{+0.026}_{-0.016}$ in $\Lambda_{\rm s}$CDM, substantially reducing the discrepancy with KiDS-1000, whereas Planck-$\Lambda$CDM gives $S_8 = 0.832 \pm 0.013$, resulting in a $3.1\sigma$ discrepancy. Just as the interplay between $\Omega_{\rm m0}$ and $S_8$ allows $\Lambda_{\rm s}$CDM to alleviate the $S_8$ tension, its inherent tendency to predict lower $\Omega_{\rm m0}$ values relative to $\Lambda$CDM in observational analyses provides a potential explanation for the suppression of structure growth reported in~\cite{Nguyen:2023fip}. 
Rather than requiring an anomalously high $\gamma \approx 0.64$, as inferred when $\gamma$ is treated as a free parameter in $\Lambda$CDM~\cite{Nguyen:2023fip}, $\Lambda_{\rm s}$CDM has the potential to achieve a similar suppression by reducing $\Omega_{\rm m}(a)$, and correspondingly $\Omega_{\rm m0}$ as well, in the late universe while maintaining $\gamma \approx 0.55$, as expected in GR.

Consequently, the $\Lambda_{\rm s}$CDM framework presents a compelling possibility for a unified explanation of the major cosmological tensions---namely, the $H_0$, $S_8$, and the recently identified $\gamma$ tensions---through a single underlying physical mechanism. In the following, we examine this potential by considering the \textit{abrupt} $\Lambda_{\rm s}$CDM scenario~\cite{Akarsu:2021fol,Akarsu:2022typ,Akarsu:2023mfb} as a proxy for rapid transitions, under the assumption that gravity is governed by GR.

\section{Methodology and data}
\label{sec:MandD}

In this work, we focus on two models: $\gamma\Lambda$CDM and $\gamma\Lambda_{\rm s}$CDM, which extend $\Lambda$CDM and abrupt $\Lambda_{\rm s}$CDM by treating $\gamma$ as a free parameter, yielding seven and eight free parameters, respectively. We stress again that $\gamma$ is treated here as an effective, phenomenological diagnostic of the linear growth sector,
rather than as a fundamental parameter of the standard model $\Lambda$CDM or of the abrupt $\Lambda_{\rm s}$CDM scenario (see Ref.~\cite{Akarsu:2025ijk} for a GR-based perturbation analysis of this model). Both models share seven parameters, six of which are standard in $\Lambda$CDM: the physical baryon density $\Omega_{\rm b} h^2$, the physical cold dark matter density $\Omega_{\rm c} h^2$, the optical depth $\tau$, the angular scale of the sound horizon at recombination $\theta_s$, the amplitude of primordial scalar perturbations $\log(10^{10} A_s)$, and the scalar spectral index $n_s$. 
The seventh shared parameter is the growth index $\gamma$, which governs the evolution of matter perturbations. The distinguishing feature of $\gamma\Lambda_{\rm s}$CDM is an additional parameter, $z_\dagger$, which sets the redshift of the mirror AdS-to-dS transition. We consider the extension of the abrupt $\Lambda_{\rm s}$CDM framework~\eqref{eqn:abruptlscdm} for simplicity, serving as a phenomenologically idealized proxy for a sufficiently rapid yet smooth mirror AdS-to-dS transition. 
The priors applied to each free parameter are summarized in~\cref{table:priors}.

\begin{table}[!t]
\caption{Ranges for the flat prior distributions imposed on cosmological parameters.}
	\begin{center}
		\footnotesize
		\renewcommand{\arraystretch}{1.1}
		\begin{tabular}{l@{\hspace{0. cm}}@{\hspace{1.5 cm}} c}
			\hline\hline
			\textbf{Parameter} & \textbf{Prior} \\
			\hline\hline
			$\Omega_{\rm b} h^2$ & $[0.005\,,\,0.1]$ \\
			$\Omega_{\rm c} h^2$ & $[0.005\,,\,0.99]$ \\
			$\tau$ & $[0.01, 0.8]$ \\
			$100\,\theta_s$ & $[0.5\,,\,10]$ \\
			$\log(10^{10}A_{\rm s})$ & $[1.61\,,\,3.91]$ \\
			$n_{\rm s}$ & $[0.8\,,\, 1.2]$ \\
			$\gamma$ & $[0.0\,,\,2.0]$ \\
                $z_\dagger$ & $[1.0\,,\,3.0]$ \\
			\hline\hline
		\end{tabular}
		\label{table:priors}
	\end{center}
\end{table}

Given that the primary goal of this study is to compare the $\gamma\Lambda$CDM and $\gamma\Lambda_{\rm s}$CDM models, particularly in terms of the inferred value of $\gamma$ relative to the theoretical prediction $\gamma\approx 0.55$, a rigorous parameter inference analysis is essential. To accomplish this, we utilize \textsc{Cobaya}~\cite{Torrado:2020dgo}, a Markov chain Monte Carlo (MCMC) sampler designed for cosmological models, in conjunction with \textsc{CAMB}~\cite{Lewis:1999bs}, a Boltzmann solver.  
Specifically, we use a modified version called \textsc{CAMB GammaPrime Growth}~\cite{minh_code} to account for the parametrization of the growth rate. The modification is introduced at the level of the late-time \emph{linear growth history} through the growth-index ansatz
$f(a)\equiv d\ln D/d\ln a=\Omega_m(a)^{\gamma}$, from which we obtain the linear growth factor $D(a)$ and evolve the linear matter power spectrum accordingly. As a result, late-time clustering quantities such as $\sigma_8(z)$ and $f\sigma_8(z)$ are modified consistently. Following the approach of Ref.~\cite{Nguyen:2023fip}, and since the growth-index parametrization is validated for subhorizon perturbations, we isolate the effect of $\gamma$ from the \emph{unlensed} primary CMB anisotropies; the modification enters the CMB only through the lensing potential (and hence lensed spectra/lensing reconstruction, when included). To evaluate the convergence of our chains, we use the Gelman-Rubin parameter $R-1$~\cite{Gelman:1992zz}, adopting $R-1<0.01$ as the threshold value for convergence.  

Alongside the $\gamma\Lambda$CDM and $\gamma\Lambda_{\rm s}$CDM models, we also consider two special cases of $\gamma\Lambda_{\rm s}$CDM in which the transition redshift is fixed at $ z_{\dagger} = 1.7 $ and $2.16$, effectively reducing the number of free parameters by one to match that of the $\gamma\Lambda$CDM model.
The first choice is motivated by previous studies identifying $z_{\dagger}\approx 1.7$ as a ``sweet spot'' of the model, where multiple cosmological tensions (including $H_0$, $M_{\rm B}$, and $S_8$) can be jointly alleviated~\cite{Akarsu:2021fol,Akarsu:2022typ,Akarsu:2023mfb,Yadav:2024duq,Akarsu:2024eoo}. In the present work, we report the fixed $z_\dagger=1.7$ case only for the PL18 and PL18+$f\sigma_8$ combinations, enabling a direct comparison with Ref.~\cite{Nguyen:2023fip} and serving as a benchmark for how such a transition affects the inferred $\gamma$ tension in these restricted datasets. Accordingly, even though this specific transition redshift may not be directly favored by the datasets considered here, we find it insightful to explicitly examine this scenario due to its proven success in addressing major cosmological tensions. The second choice is motivated \emph{a posteriori} by the best-fit value $z_{\dagger}\approx 2.16$ obtained for the PL18+$f\sigma_8$+PP\&SH0ES combination (the case with the largest improvement in fit relative to $\gamma\Lambda$CDM); we therefore consider the fixed $z_\dagger=2.16$ run only for this combination as a stability check, removing the $z_\dagger$ degeneracy and tightening the remaining constraints.

To perform parameter inference and systematically compare the models under consideration, we utilize the following datasets:
\begin{enumerate}[label=(\roman*),nosep]
    \item The \textit{Planck} 2018 CMB temperature anisotropy and polarization power spectra, their cross-spectra, and the lensing reconstruction from the \textit{Planck} 2018 legacy data release~\cite{Planck:2018nkj} (PR3). We refer to this dataset as \textbf{PL18}.
    
    \item Measurements of the $f\sigma_8$ parameter combination from redshift-space distortions (RSD), peculiar velocity, and cosmological distance measurements~\cite{Said:2020epb,Beutler:2012mnras,Huterer:2016uyq,Boruah:2019icj,Turner:2022mla,Blake:2011mnras,Blake:2013nif,Howlett:2014opa,Okumura:2015lvp,Pezzotta:2016gbo,eBOSS:2020yzd}. We refer to this dataset as \textbf{$f\sigma_8$}.
    
    \item Baryon acoustic oscillation (BAO) and redshift-space distortion (RSD) measurements from the completed SDSS-IV/eBOSS survey~\cite{eBOSS:2020yzd}. These include isotropic and anisotropic distance and expansion rate measurements at both low and high redshifts (including Lyman-$\alpha$ BAO data). We refer to this likelihood as \textbf{SDSS}. 

    \item BAO distance measurements from the first year of observations with the Dark Energy Spectroscopic Instrument (DESI)~\cite{DESI:2024mwx}. We refer to this dataset as \textbf{DESI}. Given the risk of double counting, this dataset is not used in combination with \textbf{SDSS}.
    
    \item The catalog of 1701 light curves from 1550 distinct supernovae in the PantheonPlus SN~Ia sample, covering a redshift range of $0.01<z<2.26$~\cite{Scolnic:2021amr}. We refer to this dataset as \textbf{PP}.

    \item The SH0ES Cepheid distance-ladder calibration of the PantheonPlus SN~Ia sample~\cite{Riess:2021jrx}, i.e., the calibration of the SN~Ia absolute magnitude $M_{\rm B}$ using the SH0ES host-galaxy distances. We refer to this dataset as \textbf{PP\&SH0ES}.
    
\end{enumerate}

\section{Results and discussion}

\begin{figure}[t!]
   \centering
   \includegraphics[trim = 9mm  10mm 0mm 8mm, clip, width=\columnwidth, height=1.043956\columnwidth]{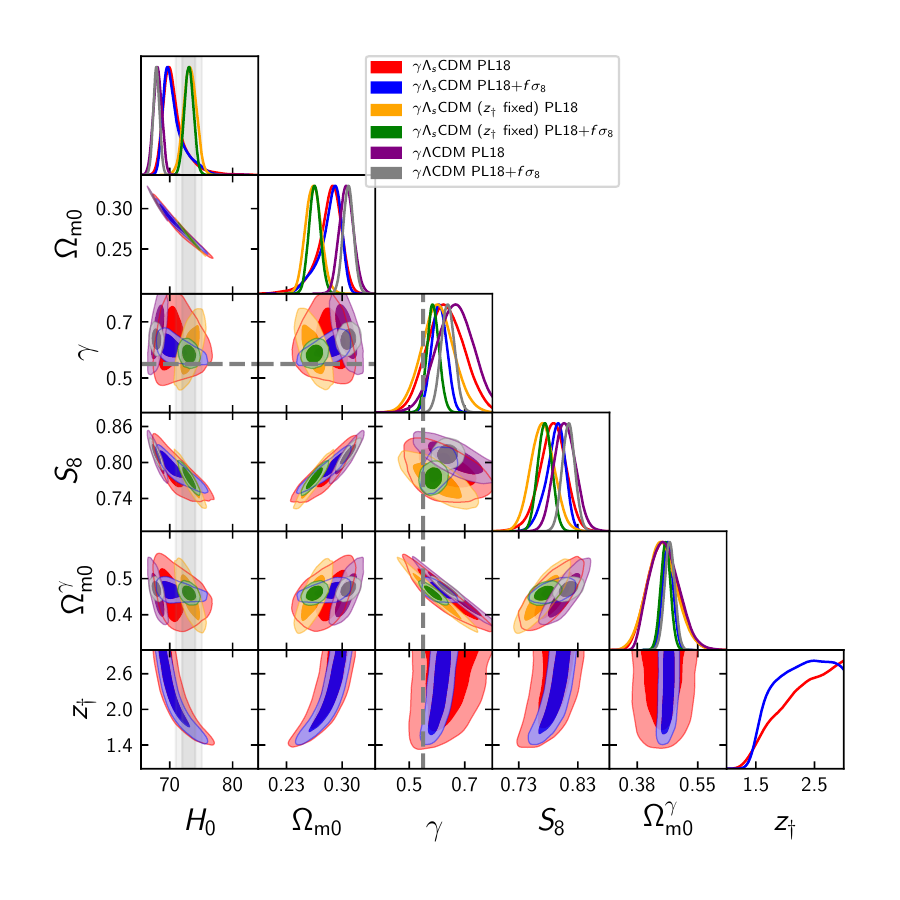}
   \caption{Triangle plot for the likelihoods PL18 and PL18+$f\sigma_8$ for the models $\gamma\Lambda$CDM, $\gamma\Lambda_{\rm s}$CDM, and $\gamma\Lambda_{\rm s}$CDM with $z_\dagger = 1.7$. The gray dotted line represents the theoretical value $\gamma = 0.55$, and the gray band represents the value $H_0 = 73.04 \pm 1.04~{\rm km\, s^{-1}\, Mpc^{-1}}$ found by the SH0ES collaboration.
   }
\label{fig:triangle_pl+fs8}
\end{figure}

\begin{figure}[ht!]
   \centering
   \includegraphics[trim = 0mm  0mm 0mm 0mm, clip, width=\columnwidth, height=1.045455\columnwidth]{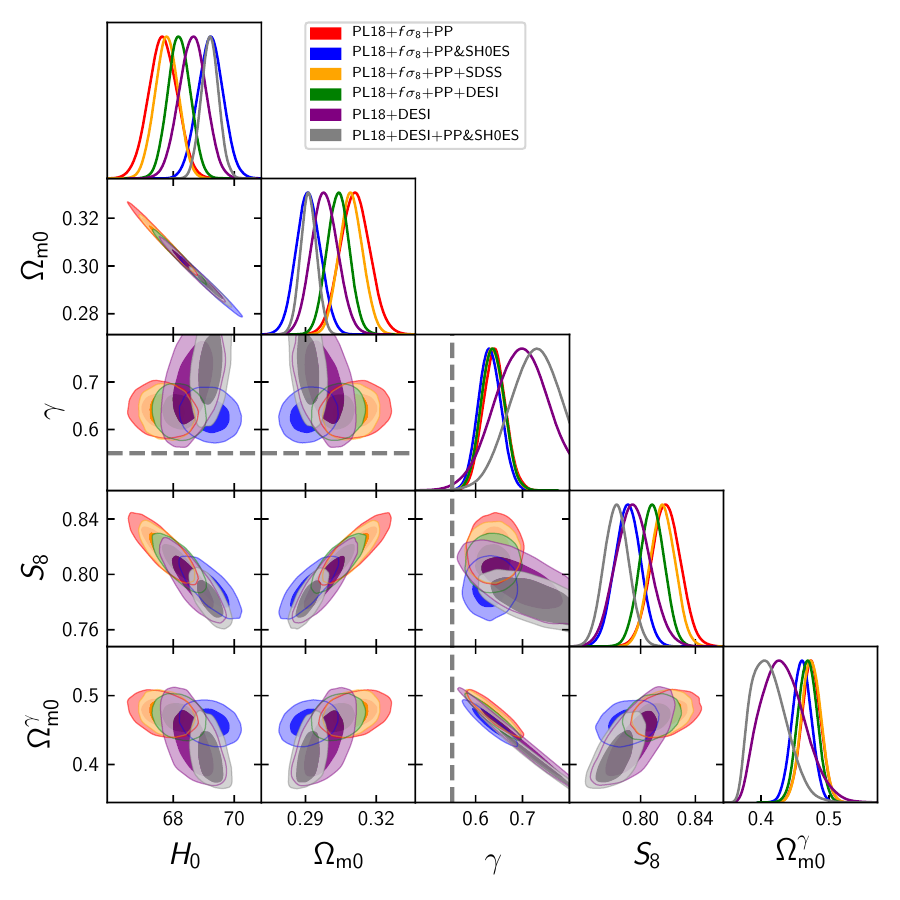}
   \caption{Marginalized posterior distributions arranged on a triangle plot for $\gamma\Lambda$CDM. The gray dotted line represents the theoretical value $\gamma = 0.55$. The only case where this value falls within the $2\sigma$ contour is when the PL18 likelihood is used alone, as seen in Fig.~\ref{fig:triangle_pl+fs8}; any other data combination is in tension with it at a level of at least $2\sigma$.}
   \label{fig:triangle_lcdm}
\end{figure}

\begin{figure}[ht!]
   \centering
   \includegraphics[trim = 0mm  0mm 0mm 0mm, clip, width=\columnwidth, height=1.045455\columnwidth]{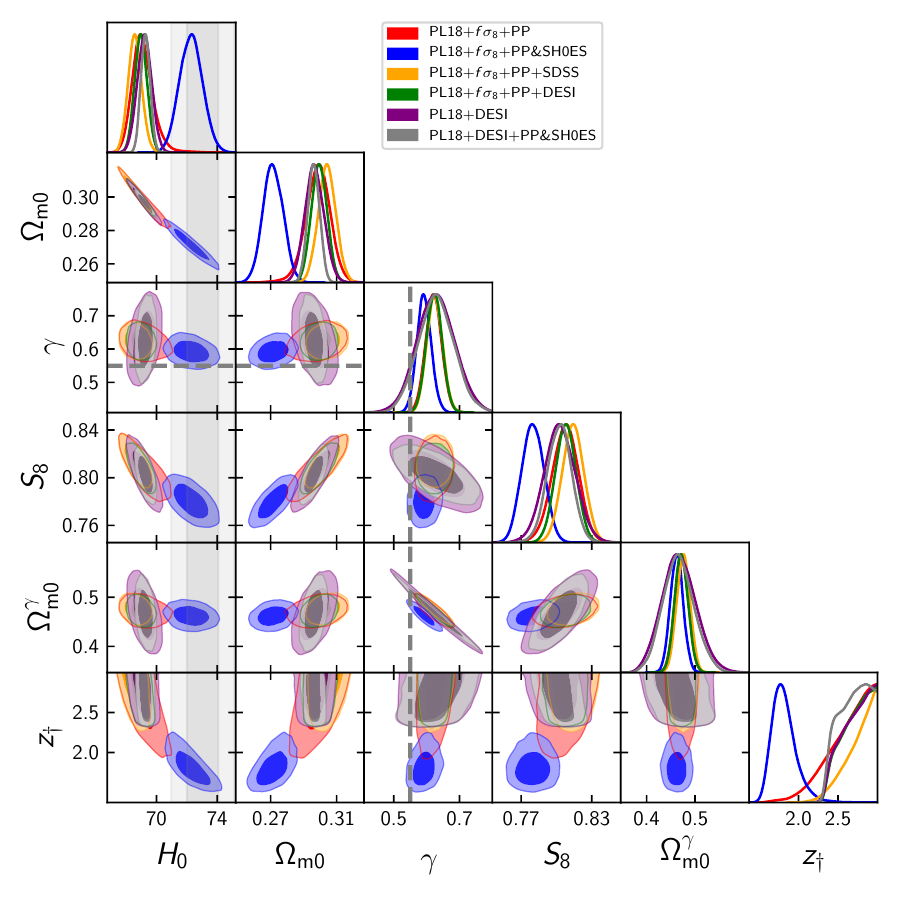}
   \caption{Marginalized posterior distributions arranged on a triangle plot for $\gamma\Lambda_{\rm s}$CDM. The gray dotted line represents the theoretical value $\gamma = 0.55$, and the gray band represents the value $H_0 = 73.04 \pm 1.04~{\rm km\, s^{-1}\, Mpc^{-1}}$ found by the SH0ES collaboration.
   }
   \label{fig:triangle_ls}
\end{figure}

\begin{figure}[ht!]
    \centering
    \includegraphics[trim = 0mm  0mm 0mm 0mm, clip, width=\columnwidth, height=1.045455\columnwidth]{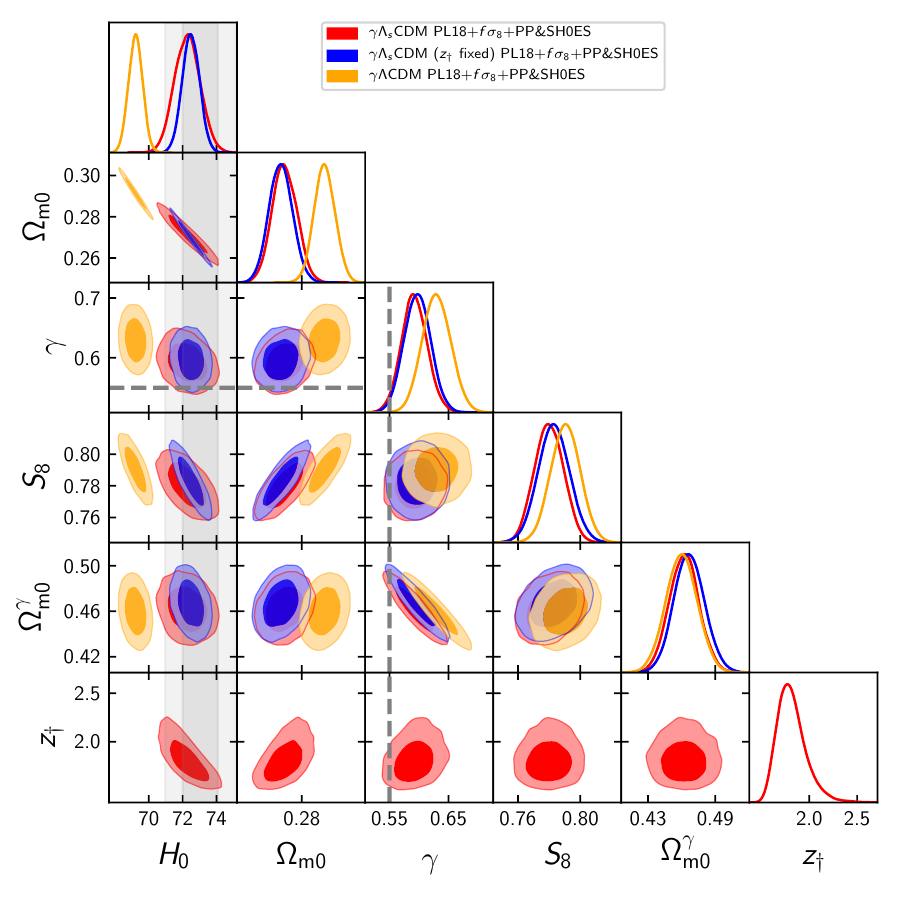}
    \caption{Triangle plot for the likelihood PL18+$f\sigma_8$+PP\&SH0ES for the models $\gamma\Lambda$CDM, $\gamma\Lambda_{\rm s}$CDM, and $\gamma\Lambda_{\rm s}$CDM with $z_\dagger=2.16$. The gray dotted line represents the theoretical value of $\gamma=0.55$, and the gray band represents the value of $H_0=73.04\pm1.04~{\rm km\, s^{-1}\, Mpc^{-1}}$ found by the SH0ES collaboration.
    }
    \label{fig:triangle_pl+fs8+pp+sh0es_fixed}
 \end{figure}

In this section, we present and interpret our parameter-inference results for the two main models---$\gamma\Lambda$CDM and $\gamma\Lambda_{\rm s}$CDM---as well as two special fixed-$z_\dagger$ cases of $\gamma\Lambda_{\rm s}$CDM.
We assess how each model addresses the growth-index tension by comparing the inferred $\gamma$ to its theoretically expected value of $\approx 0.55$, while simultaneously examining the implications for the $H_0$ and $S_8$ tensions. We also pay special attention to the constraints on the present-day growth rate $f_0 \equiv f(z=0)\simeq \Omega_{\rm m0}^\gamma$, highlighting its role in distinguishing how $\gamma\Lambda$CDM and $\gamma\Lambda_{\rm s}$CDM achieve late-time growth suppression. We report $f_0$ in Table~\ref{tab:results} as a derived diagnostic within the growth-index parametrization, while the RSD likelihood is implemented in terms of $f\sigma_8(z)$.
We present the constraints on the model parameters and relative $\chi^2_{\rm min}$ values in~\cref{tab:results}, along with the corresponding 1D and 2D posterior distributions arranged in triangle contour plots in~\cref{fig:triangle_pl+fs8,fig:triangle_lcdm,fig:triangle_ls,fig:triangle_pl+fs8+pp+sh0es_fixed}. The corresponding tension significances\footnote{\label{footnote:tail} Tension significances are computed directly from the full posterior. 
For a parameter $x$ with posterior $p(x)$ and a reference value $x_{\rm ref}$, we evaluate the two-sided tail probability
$p_{\rm tail}=2\,\min\!\left[\int_{-\infty}^{x_{\rm ref}}p(x)\,dx,\ \int_{x_{\rm ref}}^{\infty}p(x)\,dx\right]$,
and report the equivalent Gaussian significance
$N_\sigma=\Phi^{-1}\!\left(1-\frac{p_{\rm tail}}{2}\right)$,
where $\Phi^{-1}$ denotes the inverse cumulative distribution function of the standard normal distribution.}
for $\gamma$, $H_0$, and $M_{\rm B}$ are summarized in~\cref{tab:results2}.

\begin{table*}[htbp]
  \caption{Constraints on the growth index $\gamma$ and selected cosmological parameters from extended combinations of datasets. The derived present-day growth rate $f_0 \equiv \Omega_{\rm m0}^{\gamma}$ (within the growth-index parametrization) is also reported as a diagnostic quantity. The reported errors correspond to the 68\% confidence level, except for $z_\dagger$, which is given at the 95\% confidence level. The quantity $\Delta\chi^2_{\rm min}$ indicates the difference in the fit of the data between the alternative model $\gamma\Lambda_{\rm s}$CDM and $\gamma\Lambda$CDM for a given likelihood. If $\Delta\chi^2_{\rm min} < 0$ ($> 0$), then $\gamma\Lambda_{\rm s}$CDM provides a better (worse) fit to the data.}%
  \label{tab:results}%
  \centering%
  \resizebox{\textwidth}{!}{%
    \begin{tabular}{clccccccccc}%
      \toprule%
      Model & Data & $z_\dagger$ & $\gamma$ & $S_8$ & $H_0$ & $\Omega_{\rm m0}$ & $f_0 \equiv \Omega_{\rm m0}^{\gamma}$ &  $M_{\rm B}$ & $\Delta\chi^2_{\rm min}$ \\%
      \midrule%
      \multirow{8}{*}{\shortstack[l]{$\gamma\Lambda_{\rm s}$CDM($z_\dagger$ free)}}%
         & PL18 & $>1.58$ & $0.631\pm0.065$  & $0.792\pm0.021$ & $70.76^{+0.72}_{-2.1}$ & $0.283^{+0.018}_{-0.010}$ & $0.453^{+0.039}_{-0.045}$ & $\cdots$ & $1.33$ \\%
         & PL18+$f\sigma_8$ & $>1.58$ & $0.607\pm0.027$ & $0.791\pm0.007$ & $70.67^{+0.72}_{-2.0}$ & $0.284^{+0.017}_{-0.008}$ & $0.466\pm0.016$ & $\cdots$ & $-0.12$ \\%
         & PL18+$f\sigma_8$+PP & $>2.12$ & $0.621\pm0.024$ & $0.807\pm0.011$ & $69.12^{+0.53}_{-0.77}$ & $0.298^{+0.008}_{-0.007}$ & $0.472\pm0.015$ &  $-19.364^{+0.017}_{-0.024}$ & $1.85$ \\%
         & PL18+$f\sigma_8$+PP\&SH0ES & $1.82^{+0.32}_{-0.28}$ & $0.592\pm 0.022$ & $0.780\pm 0.009$ & $72.28\pm 0.76$ & $0.271\pm 0.006$ & $0.462\pm 0.013$ &  $-19.267\pm 0.023$ & $-27.36$ \\%
         & PL18+$f\sigma_8$+PP+SDSS & $>2.29$ & $0.624\pm0.024$ & $0.813\pm0.007$ & $68.66\pm0.44$ & $0.303\pm0.005$ & $0.475\pm0.014$  &  $-19.381\pm 0.014$ & $3.21$ \\%
         & PL18+$f\sigma_8$+PP+DESI & $>2.40$ & $0.623\pm0.024$ & $0.801\pm0.007$ & $68.98\pm0.41$ & $0.298\pm0.005$ & $0.472\pm0.014$ &  $-19.368\pm 0.013$ & $-1.18$ \\%
         & PL18+DESI & $>2.40$ & $0.628\pm0.057$ & $0.803\pm0.013$ & $69.21\pm0.48$ & $0.297\pm0.006$ & $0.467\pm0.035$ & $\cdots$ & $-0.69$ \\%
         & PL18+DESI+PP\&SH0ES & $> 2.38$ & $0.629\pm 0.054$ & $0.803\pm0.011$ & $69.28\pm 0.30$ & $0.296\pm 0.004$ & $0.466^{+0.030}_{-0.034}$ &  $-19.359\pm 0.009$ & $-11.98$ \\%
      \midrule%
      \multirow{2}{*}{\shortstack[l]{$\gamma\Lambda_{\rm s}$CDM($z_\dagger=1.7$)}}%
         & PL18 & {[1.7]} & $0.602\pm0.059$ & $0.771\pm0.014$ & $73.20\pm1.00$  & $0.264\pm0.011$ & $0.448\pm0.036$  & $\cdots$ & $4.41$ \\%
         & PL18+$f\sigma_8$ & {[1.7]} & $0.586\pm0.022$ & $0.774\pm0.011$ & $73.06\pm0.73$ & $0.265\pm0.007$ & $0.461\pm0.015$  & $\cdots$ & $1.97$ \\%
     \midrule%
   $\gamma\Lambda_{\rm s}$CDM($z_\dagger=2.16$)%
       & PL18+$f\sigma_8$+PP\&SH0ES & {[2.16]} & $0.591\pm0.021$ & $0.781\pm0.009$ & $72.29\pm0.71$  & $0.276\pm0.005$ & $0.467\pm0.011$ & $-19.267\pm 0.021$ & $-27.48$ \\
      \midrule%
      \multirow{8}{*}{\shortstack[l]{$\gamma\Lambda$CDM}}%
         & PL18 & $\cdots$ & $0.662\pm0.063$ & $0.808\pm0.012$ & $68.07\pm0.68$ & $0.305\pm0.009$ & $0.456\pm0.035$ & $\cdots$ & $\cdots$ \\%
         & PL18+$f\sigma_8$ & $\cdots$ & $0.639\pm0.025$ & $0.814\pm0.009$ & $67.89\pm0.50$ & $0.308\pm0.007$ & $0.471\pm0.015$ & $\cdots$ & $\cdots$ \\%
         & PL18+$f\sigma_8$+PP & $\cdots$ & $0.640\pm0.024$ & $0.818\pm0.008$ & $67.66\pm0.47$ & $0.311\pm0.006$ & $0.473\pm0.014$ &  $-19.410\pm 0.015$ & $\cdots$ \\%
         & PL18+$f\sigma_8$+PP\&SH0ES & $\cdots$ & $0.630\pm 0.024$ & $0.807\pm0.008$ & $69.22\pm 0.42$ & $0.291\pm 0.005$ & $0.460\pm 0.014$ &  $-19.361\pm 0.013$ & $\cdots$ \\%
         & PL18+$f\sigma_8$+PP+SDSS & $\cdots$ & $0.636\pm0.025$ & $0.815\pm0.007$ & $67.82\pm0.39$ & $0.309\pm0.005$ & $0.473\pm0.014$ & $-19.406\pm 0.012$ & $\cdots$ \\%
         & PL18+$f\sigma_8$+PP+DESI & $\cdots$ & $0.637\pm0.025$ & $0.808\pm0.006$ & $68.19\pm0.37$ & $0.304\pm0.005$ & $0.468\pm0.014$ & $-19.393\pm 0.012$ & $\cdots$ \\%
         & PL18+DESI & $\cdots$ & $0.693^{+0.060}_{-0.051}$ & $0.795\pm0.007$ & $68.63\pm0.44$ & $0.298\pm0.006$ & $0.432\pm0.029$ & $\cdots$ & $\cdots$ \\%
         & PL18+DESI+PP\&SH0ES & $\cdots$ & $0.719^{+0.060}_{-0.037}$ & $0.791\pm0.007$ & $69.20\pm 0.28$ & $0.295\pm 0.003$ & $0.413^{+0.018}_{-0.032}$ & $-19.361\pm 0.009$ & $\cdots$ \\%
      \bottomrule%
    \end{tabular}%
  }%
\end{table*}

\begin{table}[t!]
  \caption{Significance of the $H_0$, $\gamma$, and $M_{\rm B}$ tensions across extended combinations of datasets. Tensions are quoted as equivalent Gaussian significances computed from the two-sided posterior tail probability (see~footnote~\ref{footnote:tail}). For $\gamma$, we take the GR+$\Lambda$CDM reference value $\gamma_{\rm ref}=0.55$. For $H_0$, we compare to the SH0ES measurement $H_0^{\rm SH0ES}=73.04\pm1.04~{\rm km\,s^{-1}\,Mpc^{-1}}$. For $M_{\rm B}$, we compare to the SH0ES calibration $M_B^{\rm SH0ES}=-19.244\pm0.037$ mag~\cite{Camarena:2021jlr}.}
  \label{tab:results2}
  \centering
  \begin{ruledtabular}
    \begin{tabular}{clccc}
      Model & Data & $\sigma_\gamma$ & $\sigma_{H_0}$ & $\sigma_{M_B}$ \\
      \midrule
      \multirow{8}{*}{\shortstack[c]{$\gamma\Lambda_{\rm s}$CDM\\ 
      ($z_\dagger$ free)}}
         & PL18                              & $1.2\sigma$ & $1.1\sigma$  & $\cdots$ \\
         & PL18+$f\sigma_8$                  & $2.1\sigma$ & $1.1\sigma$  & $\cdots$ \\
         & PL18+$f\sigma_8$+PP               & $3.0\sigma$ & $2.7\sigma$  & $2.6\sigma$ \\
         & PL18+$f\sigma_8$+PP\&SH0ES        & $1.8\sigma$ & $0.6\sigma$  & $0.5\sigma$ \\
         & PL18+$f\sigma_8$+PP+SDSS          & $3.0\sigma$ & $3.9\sigma$  & $3.7\sigma$ \\
         & PL18+$f\sigma_8$+PP+DESI          & $2.8\sigma$ & $3.5\sigma$  & $3.3\sigma$ \\
         & PL18+DESI                         & $1.3\sigma$ & $3.3\sigma$  & $\cdots$ \\
         & PL18+DESI+PP\&SH0ES               & $1.4\sigma$ & $3.3\sigma$  & $2.9\sigma$ \\
      \midrule
      \multirow{2}{*}{\shortstack[c]{$\gamma\Lambda_{\rm s}$CDM\\
      ($z_\dagger=1.7$)}}
         & PL18                              & $0.8\sigma$ & $0.1\sigma$  & $\cdots$ \\
         & PL18+$f\sigma_8$                  & $1.6\sigma$ & $0.0\sigma$ & $\cdots$ \\
      \midrule
       \multirow{2}{*}{\shortstack[c]{$\gamma\Lambda_{\rm s}$CDM\\
      ($z_\dagger=2.16$)}}
         & PL18+$f\sigma_8$+PP\&SH0ES        & $1.8\sigma$ & $0.8\sigma$ & $0.5\sigma$ \\
         \\
      \midrule
      \multirow{8}{*}{\shortstack[c]{$\gamma\Lambda$CDM}}
         & PL18                              & $1.7\sigma$ & $4.1\sigma$  & $\cdots$ \\
         & PL18+$f\sigma_8$                  & $3.7\sigma$ & $4.2\sigma$  & $\cdots$ \\
         & PL18+$f\sigma_8$+PP               & $3.4\sigma$ & $4.4\sigma$  & $4.2\sigma$ \\
         & PL18+$f\sigma_8$+PP\&SH0ES        & $3.3\sigma$ & $3.3\sigma$  & $3.1\sigma$ \\
         & PL18+$f\sigma_8$+PP+SDSS          & $3.4\sigma$ & $4.4\sigma$  & $4.2\sigma$ \\
         & PL18+$f\sigma_8$+PP+DESI          & $3.2\sigma$ & $4.4\sigma$  & $4.2\sigma$ \\
         & PL18+DESI                         & $2.6\sigma$ & $3.8\sigma$  & $\cdots$ \\
         & PL18+DESI+PP\&SH0ES               & $3.2\sigma$ & $3.5\sigma$  & $3.3\sigma$ \\
    \end{tabular}
  \end{ruledtabular}
\end{table}

We observe that the $\gamma\Lambda$CDM extension essentially inherits the well-known $H_0$ tension of baseline $\Lambda$CDM, as seen in~\cref{tab:results,tab:results2} and~\cref{fig:triangle_pl+fs8}.
Across all dataset combinations---PL18, PL18+$f\sigma_8$, or even when including SN~Ia/BAO---the $H_0$ tension remains at the $3.3\text{--}4.4\sigma$ level. Simultaneously, the $\gamma$ tension often exceeds $3\sigma$, reaching its lowest value ($1.7\sigma$) in the Planck-alone case, where $\gamma=0.662\pm0.063$; this lower significance is primarily due to the large uncertainties. The tension reaches its highest value, $3.7\sigma$, for the PL18+$f\sigma_8$ combination, yielding $\gamma=0.639\pm0.025$---consistent with Ref.~\cite{Nguyen:2023fip} for the same dataset combination. 
When SN~Ia data are included, the significance decreases to $3.4\sigma$ for PL18+$f\sigma_8$+PP and to $3.3\sigma$ for PL18+$f\sigma_8$+PP\&SH0ES, with $\gamma=0.640\pm0.024$ and $\gamma=0.630\pm0.024$, respectively; further including BAO data yields comparable values ($3.4\sigma$ with SDSS and $3.2\sigma$ with DESI). Furthermore, we observe that when the $f\sigma_8$ likelihood is involved in the analysis, the inferred value of $\gamma$ across different datasets consistently yields $\gamma\sim 0.638$, as seen in~\cref{tab:results} and~\cref{fig:triangle_pl+fs8,fig:triangle_lcdm}. In the analyses that do not include $f\sigma_8$ data, we observe relatively higher mean values of $\gamma$, with this effect being particularly pronounced for the PL18+DESI and PL18+DESI+PP\&SH0ES combinations (see~\cref{fig:triangle_lcdm}). 
However, since the uncertainties on $\gamma$ in these cases do not improve significantly compared to the Planck-alone analysis, the $\gamma$ tension worsens relative to the PL18 case but remains at $2.6\text{--}3.2\sigma$, in contrast to dataset combinations that include the $f\sigma_8$ likelihood, where the tension consistently exceeds $3\sigma$. 
Correspondingly, $H_0$ and $\Omega_{\rm m0}$ remain close to their Planck-alone predictions---$H_0=68.07\pm0.68\,\mathrm{km\,s^{-1}\,Mpc^{-1}}$ and $\Omega_{\rm m0}=0.305\pm0.009$---at the $1\text{--}2\sigma$ level. These values are also closely aligned with the standard Planck-$\Lambda$CDM model~\cite{Planck:2018vyg}.

We see that the present-day growth rate, $f_0 = \Omega_{\rm m0}^\gamma$, remains largely unchanged across all dataset combinations for $\gamma\Lambda$CDM and closely matches that of $\gamma\Lambda_{\rm s}$CDM in most cases.
Notable exceptions arise for PL18+DESI and PL18+DESI+PP\&SH0ES, where $\gamma\Lambda$CDM yields modestly lower $f_0$ values. This suggests that in $\gamma\Lambda$CDM, the late-time growth suppression is achieved primarily by increasing $\gamma$ beyond its GR-predicted value of $0.55$, rather than by lowering $\Omega_{\rm m0}$. Consequently, in $\gamma\Lambda$CDM, the $H_0$ value remains anchored near the Planck-likelihood baseline, $H_0=68.07\pm0.68\,\mathrm{km\,s^{-1}\,Mpc^{-1}}$, failing to reconcile the SH0ES measurement of $H_0=73.04\pm1.04\,\mathrm{km\,s^{-1}\,Mpc^{-1}}$~\cite{Riess:2021jrx}.

 The $\gamma\Lambda_{\rm s}$CDM model addresses the late-time growth suppression by lowering $\Omega_{\rm m0}$ (and consequently increasing $H_0$), rather than elevating the growth-index $\gamma$ to values around $0.64$, as is clearly seen in the $\Omega_{\rm m0}$--$\gamma$ contours of~\cref{fig:triangle_pl+fs8}. We observe from~\cref{tab:results} that $\Omega_{\rm m0}$ can be lower in $\gamma\Lambda_{\rm s}$CDM than in $\gamma\Lambda$CDM when the data allow a transition redshift $z_\dagger\lesssim 2$, whereas for combinations including SDSS or DESI BAO data---which push $z_\dagger\gtrsim 2.3$ and $\gtrsim 2.4$, respectively---the shift in $\Omega_{\rm m0}$ becomes small and can be negligible (e.g., PL18+DESI+PP\&SH0ES; \cref{tab:results}). Despite this difference, the present-day growth rate, $f_0 = \Omega_{\rm m0}^\gamma$, remains nearly identical in both models in most cases, meaning the final suppression of structure growth is comparable---but crucially, $\gamma\Lambda_{\rm s}$CDM achieves this without requiring $\gamma$ to significantly exceed 0.6. As a result,~\cref{tab:results2} shows that $\gamma\Lambda_{\rm s}$CDM generally reduces both the $\gamma$ and $H_0$ tensions relative to $\gamma\Lambda$CDM---often while predicting $S_8\sim0.80$ values lower than the Planck-$\Lambda$CDM expectation ($S_8=0.832\pm0.013$~\cite{Planck:2018vyg}) and consistent with Planck-$\Lambda_{\rm s}$CDM ($S_8=0.801^{+0.026}_{-0.016}$~\cite{Akarsu:2023mfb,Akarsu:2024eoo})---bringing the tensions, especially in the low-$z_\dagger$ cases, to the $\sim1\text{--}3\sigma$ level, and in the most favorable combinations even below $2\sigma$, rather than leaving the $H_0$ tension at $3.3\text{--}4.4\sigma$ and the $\gamma$ tension typically at the $\sim3\sigma$ level or above in $\gamma\Lambda$CDM. The close agreement in $f_0$ between the two models, despite their stark differences in $\Omega_{\rm m0}$ and $\gamma$, aligns directly with the mechanism outlined in~\cref{sec:mechanism}, which was anticipated to simultaneously address the $\gamma$, $H_0$, and $S_8$ tensions.

As shown in~\cref{tab:results2}, the $\gamma\Lambda_{\rm s}$CDM model exhibits good performance in alleviating the $H_0$ tension, mirroring the standard ($\gamma=0.55$) $\Lambda_{\rm s}$CDM framework~\cite{Akarsu:2021fol,Akarsu:2022typ,Akarsu:2023mfb,Yadav:2024duq,Akarsu:2024eoo}. In particular, for the Planck-alone and PL18+$f\sigma_8$ analyses, both of which yield a lower bound of $z_\dagger > 1.6$ for the mirror AdS-to-dS transition redshift, the $H_0$ tension is reduced to a mild $1.1\sigma$ discrepancy, with values of $H_0=70.76^{+0.72}_{-2.1}\,\mathrm{km\,s^{-1}\,Mpc^{-1}}$ and $H_0=70.67^{+0.72}_{-2.0}\,\mathrm{km\,s^{-1}\,Mpc^{-1}}$, respectively, though we note the relatively larger uncertainties and non-Gaussianities in the posteriors and refer the reader to \cref{fig:triangle_pl+fs8} for the marginalized posteriors. The improvement in the $\gamma$ tension follows a similar trend, with the Planck-alone analysis yielding $\gamma=0.631\pm0.065$, corresponding to a $1.2\sigma$ tension with the theoretically expected value, while the PL18+$f\sigma_8$ combination gives $\gamma=0.607\pm0.027$, corresponding to a $2.1\sigma$ tension. 
We emphasize that in the Planck-alone case, the relatively larger uncertainties contribute to the reduced significance of the $\gamma$ tension. Additionally, we note that $S_8\approx0.79$ remains consistent with the predictions of the standard Planck-$\Lambda_{\rm s}$CDM and KiDS-$\Lambda_{\rm s}$CDM analyses~\cite{Akarsu:2023mfb,Akarsu:2024eoo}, implying that our results do not indicate deviations from the standard $\Lambda_{\rm s}$CDM framework. Incorporating SN~Ia data further tightens the constraints and pushes the transition to a higher redshift. In the PL18+$f\sigma_8$+PP case, the lower bound on the transition redshift increases slightly to $z_\dagger > 2.12$, diminishing the impact of the mirror AdS-to-dS transition. 
Consequently, the $H_0$ and $\gamma$ tensions increase to $2.7\sigma$ and $3.0\sigma$, respectively, but this still constitutes a non-negligible improvement over $\gamma\Lambda$CDM, which exhibits a $3.4\sigma$ $\gamma$ tension and a $4.4\sigma$ $H_0$ tension for the same dataset combination. 
In contrast, for the PL18+$f\sigma_8$+PP\&SH0ES combination, the preferred transition redshift remains in the low-$z$ regime, $z_\dagger=1.82^{+0.32}_{-0.28}$, allowing the model to operate more effectively. This results in a reduction of both the $\gamma$ and $H_0$ tensions, bringing them below the commonly considered $2.0\sigma$ threshold for a tension to be statistically significant, down to $1.8\sigma$ and $0.6\sigma$, respectively, whereas both tensions are at the $3.3\sigma$ level in the $\gamma\Lambda$CDM model. 
The same trend is reflected in the calibration-based $M_{\rm B}$ tension, which is reduced to $0.5\sigma$ in $\gamma\Lambda_{\rm s}$CDM, compared with $3.1\sigma$ in $\gamma\Lambda$CDM. Moreover, the corresponding value, $S_8=0.780\pm0.009$, remains compatible with the standard $\Lambda_{\rm s}$CDM analyses based on Planck and KiDS-1000~\cite{Akarsu:2023mfb,Akarsu:2024eoo}.

We note that the derived present-day growth rate $f_0=\Omega_{\rm m0}^{\gamma}$ in $\gamma\Lambda_{\rm s}$CDM is mutually consistent across the dataset combinations considered, in most cases well within $1\sigma$ (Table~\ref{tab:results}). The same is largely true in $\gamma\Lambda$CDM for PL18 and for combinations including the external $f\sigma_8$ likelihood, for which the inferred $f_0$ values are consistent with those of $\gamma\Lambda_{\rm s}$CDM. The exceptions are the combinations that include DESI but not the external $f\sigma_8$ likelihood (PL18+DESI and PL18+DESI+PP\&SH0ES), where $\gamma\Lambda$CDM prefers lower $f_0$ values. 
However, even in the most pronounced case (PL18+DESI+PP\&SH0ES, for which $\gamma\Lambda_{\rm s}$CDM provides a substantially better fit; see Table~\ref{tab:results}), the corresponding shift in $f_0$ remains at the $\lesssim2\sigma$ level relative to the $\gamma\Lambda_{\rm s}$CDM value. This consistency in the inferred values of $f_0$ leads to the positive correlation observed between $\Omega_{\rm m0}$ and $\gamma$ in the $\gamma\Lambda_{\rm s}$CDM model, as seen in the 2D contour plots of $\Omega_{\rm m0}$-$\gamma$ in~\cref{fig:triangle_pl+fs8,fig:triangle_ls,fig:triangle_pl+fs8+pp+sh0es_fixed}. Furthermore, we also observe a negative correlation between $\Omega_{\rm m0}$ and $H_0$, as well as between $H_0$ and $\gamma$, in the same figures. This highlights the key mechanism through which $\gamma\Lambda_{\rm s}$CDM mitigates both the $H_0$ and $\gamma$ tensions (see~\cref{sec:mechanism}). 
Specifically, for combinations that allow a transition redshift in the low-$z$ regime (roughly $z_\dagger\lesssim 2$), $\gamma\Lambda_{\rm s}$CDM tends to favor lower values of $\Omega_{\rm m0}$ than $\gamma\Lambda$CDM, and correspondingly higher $H_0$. In this regime, $\gamma\Lambda_{\rm s}$CDM typically yields $\Omega_{\rm m0}\sim 0.27$--0.30, whereas $\gamma\Lambda$CDM typically remains in the range $\Omega_{\rm m0}\sim 0.29$--0.31, closer to the Planck-$\Lambda$CDM value (see~\cref{tab:results}).
This trend is also consistent with Planck-only constraints in the standard (GR) versions of the models, for which $\Lambda_{\rm s}$CDM prefers a lower $\Omega_{\rm m0}=0.286^{+0.023}_{-0.010}$~\cite{Akarsu:2023mfb} than base $\Lambda$CDM, for which $\Omega_{\rm m0}=0.315\pm0.007$~\cite{Planck:2018vyg}. For combinations including BAO data from SDSS or DESI, however, the preferred transition redshift is typically pushed to $z_\dagger\gtrsim 2.3$--2.4, and the shift in $\Omega_{\rm m0}$ correspondingly becomes small or negligible. 
Notably, when $\Omega_{\rm m0}$ is shifted downward in $\gamma\Lambda_{\rm s}$CDM, it correlates with reduced $\gamma$ and $H_0$ tensions, with the most significant alleviation occurring for $\Omega_{\rm m0}$ values closer to $\sim 0.28$.

A closer examination of~\cref{fig:triangle_pl+fs8,fig:triangle_ls,fig:triangle_pl+fs8+pp+sh0es_fixed} reveals that $\Omega_{\rm m0}$, $\gamma$, and $S_8$ exhibit positive correlations with $z_\dagger$, while $H_0$ exhibits a negative correlation with $z_\dagger$. However, these effects become noticeable only for $z_\dagger \lesssim 2.5$ and significant for $z_\dagger \lesssim 2.0$. This is in full agreement with prior studies of $\Lambda_{\rm s}$CDM, which anticipate that if the mirror AdS-to-dS transition occurs too early, say for $z_\dagger > 2.5$, during the epoch when matter still dominates cosmic dynamics, the AdS-like CC lacks sufficient time and influence to induce meaningful deviations from the standard $\Lambda$CDM expansion history~\cite{Akarsu:2021fol,Akarsu:2022typ,Akarsu:2023mfb,Yadav:2024duq,Akarsu:2024eoo}. Indeed, we find that when $z_\dagger \lesssim 2.0$, the mirror AdS-to-dS transition can effectively reduce both the $\gamma$ and $H_0$ tensions to the ${\lesssim2\sigma}$ level. Conversely, for $z_\dagger \gtrsim 2.5$, the model effectively converges toward $\gamma\Lambda$CDM, compromising its ability to mitigate the $\gamma$ and $H_0$ tensions and offering only marginal improvements over $\gamma\Lambda$CDM.

This behavior is particularly evident when we extend the combined PL18+$f\sigma_8$+PP dataset by incorporating BAO data, either from SDSS or DESI. We find that PL18+$f\sigma_8$+PP+SDSS yields ${z_\dagger > 2.3}$ and PL18+$f\sigma_8$+PP+DESI gives ${z_\dagger > 2.4}$, placing both cases in the regime where the transition redshift is too high to fully resolve the tensions. As a result, as seen in~\cref{tab:results2}, $\gamma\Lambda_{\rm s}$CDM provides only modest improvements: the $\gamma$ tension remains at $3.0\sigma$ (SDSS) and $2.8\sigma$ (DESI), while the $H_0$ tension is at the $3.9\sigma$ (SDSS) and $3.5\sigma$ (DESI) levels. Nevertheless, these results indicate that $\gamma\Lambda_{\rm s}$CDM performs $\sim0.4\sigma$ better than $\gamma\Lambda$CDM in terms of the $\gamma$ tension, and $\sim0.5\sigma$ (SDSS) to $\sim0.9\sigma$ (DESI) better in terms of the $H_0$ tension.

We observe that in the case of the PL18+$f\sigma_8$+PP+DESI combination, $\gamma\Lambda_{\rm s}$CDM performs slightly better than in PL18+$f\sigma_8$+PP+SDSS in mitigating both the $\gamma$ and $H_0$ tensions. Moreover, when comparing the minimum $\chi^2_{\rm min}$ values, we find that $\gamma\Lambda_{\rm s}$CDM provides a worse fit to the PL18+$f\sigma_8$+PP+SDSS dataset than $\gamma\Lambda$CDM, yielding $\Delta\chi^2_{\rm min}=3.21$;
whereas for the PL18+$f\sigma_8$+PP+DESI dataset it fits better, with $\Delta\chi^2_{\rm min}=-1.18$ (Table~\ref{tab:results}). Motivated by this, we further examined DESI-based combinations without the external $f\sigma_8$ likelihood, namely PL18+DESI and PL18+DESI+PP\&SH0ES. In both cases, $\gamma\Lambda_{\rm s}$CDM yields a high transition-redshift bound ($z_\dagger>2.40$ and $z_\dagger>2.38$, respectively; Table~\ref{tab:results}), so the background approaches the $\gamma\Lambda$CDM limit and the residual $H_0$ discrepancy remains at the $\sim3\sigma$ level (Table~\ref{tab:results2}). For PL18+DESI, the improvement in fit is marginal ($\Delta\chi^2_{\rm min}=-0.69$) and the $\gamma$ tension is reduced to $1.3\sigma$, largely due to the relatively large uncertainty on $\gamma$. For comparison, in $\gamma\Lambda$CDM the PL18+DESI combination yields $\gamma=0.693^{+0.060}_{-0.051}$, corresponding to a $2.6\sigma$ $\gamma$ tension and a $3.8\sigma$ $H_0$ tension (Tables~\ref{tab:results} and~\ref{tab:results2}). 
By contrast, for PL18+DESI+PP\&SH0ES the improvement in fit is substantial ($\Delta\chi^2_{\rm min}=-11.98$), while $\gamma\Lambda_{\rm s}$CDM also yields a notably smaller $\gamma$ tension than $\gamma\Lambda$CDM ($1.4\sigma$ versus $3.2\sigma$; Table~\ref{tab:results2}), with a modest improvement in the $H_0$ tension ($3.3\sigma$ versus $3.5\sigma$). Notably, adding SN~Ia data (PL18+DESI+PP\&SH0ES) does not change the conclusion that the high-$z_\dagger$ bound limits the extent to which the $H_0$ discrepancy can be reduced in these DESI-based combinations. These results indicate that even when DESI pushes $z_\dagger$ to relatively high values, $\gamma\Lambda_{\rm s}$CDM can retain an advantage over $\gamma\Lambda$CDM, particularly in the growth sector.

These higher tensions, which arise when the BAO data are incorporated into our analysis, suggest that the DESI and SDSS datasets---both of which rely on 3D BAO measurements calibrated against Planck-$\Lambda$CDM as the fiducial model---limit the effectiveness of $\gamma\Lambda_{\rm s}$CDM in predicting $\gamma = 0.55$.\footnote{\label{fnote}The performance of the abrupt $\Lambda_{\rm s}$CDM model against pre-DESI BAO data, particularly the SDSS BAO dataset, has been extensively studied in the literature~\cite{Akarsu:2021fol,Akarsu:2022typ,Akarsu:2024eoo}. 
While SDSS BAO data statistically favor $z_\dagger \gtrsim 2.1$, limiting the model’s effectiveness, it has been shown that when combined with CMB data, the overall fit remains comparable to, or even slightly better than, standard $\Lambda$CDM in terms of Bayesian evidence~\cite{Akarsu:2022typ}. A detailed analysis in~\cite{Akarsu:2022typ} demonstrated that low-$z$ SDSS BAO data primarily weaken the model’s success, whereas high-$z$ BAO, viz., Ly-$\alpha$ BAO, favor a transition redshift of $z_\dagger \lesssim 2.3$. 
Recognizing that high-$z$ and low-$z$ BAO constraints yield distinctive correlations in the $H_0$-$\Omega_{\rm m0}$ plane within $\Lambda$CDM~\cite{eBOSS:2020yzd}, Ref.~\cite{Akarsu:2024eoo} performed an analysis using only the high-$z$ subset of SDSS BAO data ($z > 1$), finding that both the model’s fit and its ability to mitigate cosmological tensions improve significantly compared to using the full SDSS BAO dataset. Furthermore, Ref.~\cite{Akarsu:2023mfb} investigated the use of transverse BAO (BAOtr) (or 2D BAO), a less model-dependent alternative to the commonly used 3D BAO data (often referred to simply as BAO),
the latter requiring the adoption of a fiducial cosmology (often Planck-$\Lambda$CDM) to convert angles/redshifts into comoving separations, which could potentially introduce a bias
when analyzing scenarios beyond $\Lambda$CDM. They found that combining Planck with BAOtr provides very strong Bayesian evidence for $\Lambda_{\rm s}$CDM over $\Lambda$CDM, yielding $H_0 = 73.30^{+1.20}_{-1.00}\,\mathrm{km\,s^{-1}\,Mpc^{-1}}$ and
rendering the $H_0$ tension negligible ($\sim0.1\sigma$) in that combination. The robustness of this BAOtr-driven picture was further demonstrated in Ref.~\cite{Yadav:2024duq}, which extended the abrupt $\Lambda_{\rm s}$CDM model by allowing $N_{\rm eff}$ and $\sum m_{\nu}$ to vary freely: in combinations including BAOtr data, the mirror AdS-to-dS transition alone drives $H_0\approx73\,\mathrm{km\,s^{-1}\,Mpc^{-1}}$, with the transition redshift remaining consistent with the base-model value $z_\dagger\approx1.7$ and no significant deviation from the standard value $N_{\rm eff}=3.044$, indicating that the late-time transition can alleviate the major tensions without altering the standard prerecombination Universe. 
Additionally, a preference for lower transition redshifts in the presence of BAOtr data is not unique to the abrupt $\Lambda_{\rm s}$CDM model: the $w$XCDM model of Refs.~\cite{Gomez-Valent:2024tdb,Gomez-Valent:2024ejh}, in which the high-redshift component is not an AdS-like CC but an exotic fluid (``phantom matter'') with negative energy density and positive pressure, selects a transition to a standard DE component close to $z_t \simeq 1.5$ when BAOtr data are used. Similarly, the string-inspired $\Lambda_{\rm s}$CDM$^+$ realization~\cite{Anchordoqui:2023woo,Anchordoqui:2024gfa,Anchordoqui:2024dqc,Soriano:2025gxd}, which augments abrupt $\Lambda_{\rm s}$CDM with a theoretically predicted excess $\Delta N_{\rm eff}\simeq0.25$, yields $z_\dagger = 2.11 \pm 0.26$ from the Planck+BAOtr+PP\&SH0ES+KiDS-1000 combination. \label{footnote_2dbao} }
Even so, DESI-based combinations generally prefer $\gamma\Lambda_{\rm s}$CDM over $\gamma\Lambda$CDM in terms of the minimum $\chi^2$ (i.e., $\Delta\chi^2_{\rm min}<0$; Table~\ref{tab:results}), with the preference ranging from marginal (e.g., PL18+DESI, $\Delta\chi^2_{\rm min}=-0.69$) to substantial (PL18+DESI+PP\&SH0ES, $\Delta\chi^2_{\rm min}=-11.98$). Moreover, across DESI-based cases $\gamma\Lambda_{\rm s}$CDM yields a smaller $\gamma$ tension by $\sim0.4$--$1.8\sigma$ and a smaller $H_0$ tension by $\sim0.2$--$0.9\sigma$ relative to $\gamma\Lambda$CDM for the same dataset combinations (Table~\ref{tab:results2}).

For dataset combinations that include PP, we also report the inferred SN~Ia absolute magnitude $M_{\rm B}$, which can be used to quantify the Hubble-tension discrepancy in a calibration-based manner~\cite{Camarena:2019moy,Benevento:2020fev,Efstathiou:2021ocp}. 
From Table~\ref{tab:results2}, we find that $\gamma\Lambda_{\rm s}$CDM generally reduces the $M_{\rm B}$ tension relative to $\gamma\Lambda$CDM, when compared to the SH0ES calibration $M_B^{\rm SH0ES}=-19.244\pm0.037$~\cite{Camarena:2021jlr}. 
This is not surprising, since $M_{\rm B}$ is tightly correlated with $H_0$, so alleviating the discrepancy in one typically leads to a corresponding reduction in the other. 
Among the combinations considered here, the most limited improvement occurs for PL18+DESI+PP\&SH0ES, for which the inferred posteriors for $H_0$ and $\Omega_{\rm m0}$ remain very similar in the two models, leaving little room for a differential shift in $M_{\rm B}$.

As a next step, we fix $z_\dagger = 1.7$, motivated by previous studies on $\Lambda_{\rm s}$CDM showing that a mirror AdS-to-dS transition around this redshift effectively alleviates multiple major cosmological tensions, including those in $H_0$, $M_{\rm B}$, and $S_8$~\cite{Akarsu:2021fol,Akarsu:2022typ,Akarsu:2023mfb,Akarsu:2024eoo}. 
For instance, Ref.~\cite{Akarsu:2023mfb} finds $z_\dagger = 1.72^{+0.09}_{-0.12}$ from a combined analysis using Planck+BAOtr+PP\&SH0ES+KiDS-1000, yielding $H_0 = 73.16 \pm 0.64\,\mathrm{km\,s^{-1}\,Mpc^{-1}}$ and effectively eliminating the $H_0$ tension ($\sim0.1\sigma$) in that combination. For this reason, we find it interesting to explicitly examine the $z_\dagger = 1.7$ scenario for the Planck-alone and PL18+$f\sigma_8$ dataset combinations, as our free $z_\dagger$ analysis indicates that $z_\dagger = 1.7$ remains consistent with the lower bound $z_\dagger > 1.58$ inferred for these cases. 
Thus, for these two combinations, we conduct an additional analysis of the $\gamma\Lambda_{\rm s}$CDM model by fixing $z_\dagger = 1.7$, reducing the number of free parameters to the same seven as $\gamma\Lambda$CDM, and evaluating its impact on the inferred cosmological tensions. As seen in~\cref{tab:results,tab:results2}, this special case substantially reduces both the $H_0$ and $\gamma$ tensions in these two combinations. 
See also~\cref{fig:triangle_pl+fs8}, which compares PL18 and PL18+$f\sigma_8$ for $\gamma\Lambda$CDM, $\gamma\Lambda_{\rm s}$CDM, and the special fixed-$z_\dagger=1.7$ case. 
However, this value of $z_\dagger$ is disfavored by both the PL18 and PL18+$f\sigma_8$ datasets in terms of goodness of fit, as the $\gamma\Lambda_{\rm s}$CDM model worsens the $\chi^2_{\rm min}$ by $\sim\!4$ and $\sim\!2$, respectively, compared to $\gamma\Lambda$CDM. 
Moreover, in our analyses including BAO data, as well as in the PL18+$f\sigma_8$+PP case, this value of $z_\dagger$ is excluded at more than $2\sigma$; by contrast, it remains allowed in the PL18+$f\sigma_8$+PP\&SH0ES combination. 
Regardless, in the Planck-alone analysis, the lowered $\Omega_{\rm m0}$ implied by $z_\dagger=1.7$ raises $H_0$ enough to effectively eliminate the $H_0$ tension ($0.1\sigma$) and simultaneously keeps $\gamma$ near $0.55$, reducing the $\gamma$ tension to $0.8\sigma$ as well. 
A similar effect emerges when $f\sigma_8$ data are added: the PL18+$f\sigma_8$ combination still allows $H_0$ to remain sufficiently high ($\gtrsim 73\,\mathrm{km\,s^{-1}\,Mpc^{-1}}$) while keeping $\gamma\approx 0.58\text{--}0.60$, reducing the $\gamma$ tension to $1.6\sigma$ and driving the $H_0$ tension down to $0\sigma$. The $z_\dagger\simeq1.7$ case highlights the impact of a rapid mirror AdS-to-dS transition at this redshift: it can jointly reduce the $\gamma$ tension and significantly alleviate the $H_0$ tension, thereby indicating $z_\dagger\approx1.7$ as a phenomenological redshift of interest for simultaneous tension alleviation. 
This is consistent with earlier findings in standard ($\gamma=0.55$) abrupt $\Lambda_{\rm s}$CDM studies~\cite{Akarsu:2021fol,Akarsu:2022typ,Akarsu:2023mfb,Yadav:2024duq,Akarsu:2024eoo}, which identify this transition epoch as the regime where the major cosmological tensions---in $H_0$, $M_{\rm B}$, and $S_8$---are substantially reduced.

Finally, another special case in our analysis focuses on the PL18+$f\sigma_8$+PP\&SH0ES dataset, for which~\cref{tab:results} shows that $\gamma\Lambda_{\rm s}$CDM provides a substantially better fit than $\gamma\Lambda$CDM, with $\Delta\chi^2_{\min}=-27.36$. In this scenario, both the $\gamma$ and $H_0$ tensions are significantly reduced. Noting that the best-fit transition redshift for this combination is $z_\dagger \approx 2.16$ (see~\cref{fig:triangle_ls}), we fix $z_\dagger=2.16$ (removing one degree of freedom) and rerun the inference. This case yields $\Delta\chi^2_{\min}=-27.48$, and its comparison with $\gamma\Lambda_{\rm s}$CDM and $\gamma\Lambda$CDM is provided in~\cref{fig:triangle_pl+fs8+pp+sh0es_fixed}. 
The resulting constraints remain consistent with those found when $z_\dagger$ was free to vary, except that $H_0$ and $\Omega_{\rm m0}$ become more tightly constrained due to the removed degeneracy with $z_\dagger$ after fixing its value. Consequently, relative to $\gamma\Lambda$CDM for the same dataset combination, the $H_0$ tension is reduced to $\sim0.8\sigma$, while the $\gamma$ tension is $\sim1.8\sigma$, underscoring the advantage of $\gamma\Lambda_{\rm s}$CDM in mitigating both tensions in this best-fit scenario.

These results confirm the heuristic argument from~\cref{sec:mechanism}: in $\gamma\Lambda_{\rm s}$CDM, the AdS-like CC at $z>z_\dagger$ reduces $\Omega_{\rm m0}$, thereby achieving a similar low-redshift growth suppression (viz., $f_0\approx0.47$ found in Ref.~\cite{Nguyen:2023fip}, compared to the standard Planck-$\Lambda$CDM value of $f_0\approx0.53$)
while keeping $\gamma$ closer to its GR expectation ($\gamma \approx 0.55$) than in $\gamma\Lambda$CDM.
The key distinction is that $\gamma\Lambda_{\rm s}$CDM accomplishes this by lowering $\Omega_{\rm m0}$, which in turn can raise $H_0$ to approximately
$70\text{--}73\,\mathrm{km\,s^{-1}\,Mpc^{-1}}$ in the low-$z_\dagger$ cases, thereby mitigating both the Hubble and growth tensions simultaneously. 
In contrast, $\gamma\Lambda$CDM maintains a higher $\Omega_{\rm m0}$, necessitating an inflated $\gamma\sim 0.64$ (\cref{tab:results}) to match $f_0\approx0.47$, which leaves the $H_0$ tension at $3.3\text{--}4.4\sigma$. Furthermore, when the data allow a transition redshift in the low-$z$ regime (roughly $z_\dagger\lesssim 2$), the lower present-day matter density parameter values that drive the alleviation of the $\gamma$ and $H_0$ discrepancies in $\gamma\Lambda_{\rm s}$CDM also tend to lower $S_8$, since $S_8 \propto \sqrt{\Omega_{\rm m0}}$; for combinations including SDSS or DESI BAO data, which tend to push $z_\dagger$ to $\gtrsim 2.3$ and $\gtrsim 2.4$, respectively, the shift in $\Omega_{\rm m0}$ is reduced and so is the impact on $S_8$. This brings $S_8$ closer to its cosmic shear constraints, as found using the KiDS-1000 data for the abrupt $\Lambda_{\rm s}$CDM model~\cite{Akarsu:2023mfb,Akarsu:2024eoo}, indicating that the $S_8$ tension does not arise in this model. 
Thus, the results in~\cref{tab:results,tab:results2}, together with~\cref{fig:triangle_pl+fs8,fig:triangle_lcdm,fig:triangle_ls,fig:triangle_pl+fs8+pp+sh0es_fixed}, indicate an advantage of the mirror AdS-to-dS (sign-switching cosmological constant $\Lambda_{\rm s}$) in $\gamma\Lambda_{\rm s}$CDM over $\gamma\Lambda$CDM for addressing late-time tensions. 
By moving $\gamma$ closer to its theoretically motivated value $\gamma \approx 0.55$ while simultaneously raising $H_0$ toward local measurements, $\gamma\Lambda_{\rm s}$CDM naturally accounts for the observed growth suppression ($f_0 \simeq 0.47$) through a lower present-day matter density, $\Omega_{\rm m0}$, as reflected in the $f_0$ values reported in Table~\ref{tab:results}.

This implies that, treating $\gamma$ as a free parameter, the cosmological datasets used in this work all prefer a suppressed growth of structure compared to the theoretical predictions for both $\gamma\Lambda$CDM and $\gamma\Lambda_{\rm s}$CDM models; however, this preference, encapsulated in the deviation from the theoretical $\gamma \approx 0.55$ value, is reduced (reaching up to $1.8\sigma$ amelioration depending on the dataset combination) for $\gamma\Lambda_{\rm s}$CDM, implying the $\Lambda_{\rm s}$CDM framework has better consistency with its underlying theory compared to $\Lambda$CDM. Thus, the rapid mirror AdS-to-dS transition in the late universe, at $z_\dagger\sim2$, emerges as a mechanism that can contribute to alleviating the Hubble, $S_8$, and $\gamma$ tensions within a single, minimal extension to the standard cosmological model.

It is worth noting that the use of $f\sigma_8(z)$ as a proxy for the growth of structure may be subject to important biases. 
As discussed in Refs.~\cite{Sanchez:2020vvb,Semenaite:2021zxw,Semenaite:2022unt,Forconi:2025cwp}, $\sigma_8$ is defined on the scale $8\,h^{-1}\,{\rm Mpc}$, so variations in $h$ change the physical scale at which the fluctuation amplitude is evaluated. 
This can limit the degree to which $\sigma_8(z)$ provides a model-independent compression of the matter power spectrum amplitude, and may lead to apparent shifts in $\sigma_8(z)$ when comparing cosmologies with different expansion histories (in particular, in scenarios that aim to alleviate the $H_0$ tension). 
By contrast, a normalization based on $\sigma_{12}(z)$, the rms linear matter fluctuation amplitude in spheres of fixed comoving radius $12\,{\rm Mpc}$, avoids the $h$-dependent change of smoothing scale inherent in $\sigma_8(z)$.
Accordingly, we interpret the RSD-driven constraints with this caveat in mind, noting that residual systematics in $f\sigma_8$ could contribute to an apparent preference for low growth.

\section{Conclusions}

The recently identified growth-index ($\gamma$) tension~\cite{Nguyen:2023fip} adds another major discrepancy to the persistent challenges facing the $\Lambda$CDM model---namely, the $\sim5\sigma$ $H_0$ tension and the $\sim3\sigma$ $S_8$ tension. The $\gamma$ tension, persisting at the $3\text{--}4\sigma$ level, arises when $\gamma$ is treated as a free parameter in $\Lambda$CDM, yielding $\gamma \simeq 0.64$ and $\Omega_{\rm m0} \simeq 0.31$, with $\gamma$ significantly above the theoretically expected value $\gamma \approx 0.55$ under GR. This elevated $\gamma$ value corresponds to a suppressed growth rate of $f_0 \simeq 0.47$, compared to the expected $f_0 \simeq 0.53$ for $\gamma \approx 0.55$, further reinforcing the deviation from $\Lambda$CDM~\cite{Nguyen:2023fip}. 
The $\Lambda_{\rm s}$CDM framework, characterized by a rapid mirror AdS-to-dS transition---where the cosmological constant $\Lambda_{\rm s}$ switches sign---at redshift $z_\dagger \sim 2$, has emerged as a promising and economical extension of $\Lambda$CDM, 
with the potential to jointly alleviate major cosmological tensions, including those in $H_0$, $M_{\rm B}$, and $S_8$, while also predicting the age of the Universe and, when treated as free parameters, standard neutrino properties, all within a single theoretical framework~\cite{Akarsu:2019hmw,Akarsu:2021fol,Akarsu:2022typ,Akarsu:2023mfb,Yadav:2024duq,Akarsu:2024eoo,Akarsu:2024qsi}. Robust observational analyses indicate that $\Lambda_{\rm s}$CDM generally predicts a lower late-time matter density parameter than $\Lambda$CDM; specifically, Planck-$\Lambda_{\rm s}$CDM yields $\Omega_{\rm m0} \simeq 0.28$, which, when combined with the theoretically expected growth index $\gamma \approx 0.55$~\cite{Akarsu:2025ijk} under GR, leads to $f_0 \simeq 0.49$, in close agreement with~\cite{Nguyen:2023fip}. 
Furthermore, previous studies have shown that for $z_\dagger\approx1.7$, $\Lambda_{\rm s}$CDM resolves both the $H_0$ and $S_8$ tensions, bringing them below $\sim1\sigma$~\cite{Akarsu:2023mfb} by predicting $\Omega_{\rm m0} \approx 0.27$, which in turn yields $f_0 = 0.48$, in precise agreement with Ref.~\cite{Nguyen:2023fip}. These considerations make the $\Lambda_{\rm s}$CDM framework a well-motivated setting in which to examine whether the recently identified $\gamma$ tension can also be alleviated within the same late-time mechanism. Accordingly, we rigorously examined this potential by adopting the \textit{abrupt} $\Lambda_{\rm s}$CDM scenario as a proxy for rapid transitions, under the assumption that gravity is governed by GR. Following the methodology of Ref.~\cite{Nguyen:2023fip}, we performed a systematic parameter inference analysis, first using Planck-alone (PL18) and PL18+$f\sigma_8$ datasets, then extending our study to additional datasets---including SN~Ia (PantheonPlus with/without SH0ES) and BAO (SDSS, DESI)---to assess the robustness of $\gamma\Lambda_{\rm s}$CDM across different cosmological probes.

Our analysis supports the view that the $\Lambda_{\rm s}$CDM framework provides a late-time mechanism that can jointly alleviate the $\gamma$, $H_0$, and $S_8$ tensions, with an efficacy that is dataset dependent through the inferred transition redshift $z_\dagger$. A key observation is that, despite their fundamental differences, $\gamma\Lambda_{\rm s}$CDM and $\gamma\Lambda$CDM yield nearly identical present-day growth rates, $f_0 \equiv f(z=0)\simeq \Omega_{\rm m0}^\gamma$, across most of the dataset combinations considered. However, they achieve this through distinct mechanisms: $\gamma\Lambda$CDM suppresses structure growth primarily by increasing $\gamma$ above its GR expectation while keeping $\Omega_{\rm m0}$ close to the Planck-$\Lambda$CDM value, thereby exacerbating the $\gamma$ tension. 
In contrast, when the data allow a transition in the low-$z$ regime (roughly $z_\dagger\lesssim 2$), $\gamma\Lambda_{\rm s}$CDM can accommodate the same low-growth preference with $\gamma$ closer to $\sim0.55$ by lowering $\Omega_{\rm m0}$ (and correspondingly raising $H_0$), thereby reducing both the $\gamma$ and $H_0$ discrepancies; when $z_\dagger$ is pushed to higher values, as often happens in combinations including SDSS or DESI BAO data, these shifts are correspondingly reduced. This distinction is evident in the $\Omega_{\rm m0}$--$\gamma$ contours, where $\gamma\Lambda_{\rm s}$CDM exhibits a strong positive correlation, setting it apart from $\gamma\Lambda$CDM. 
Moreover, in the cases where $\Omega_{\rm m0}$ is shifted downward in $\gamma\Lambda_{\rm s}$CDM, this also tends to lower $S_8$ (since $S_8\propto\sqrt{\Omega_{\rm m0}}$), improving agreement with cosmic shear constraints such as the KiDS-1000 $\Lambda_{\rm s}$CDM estimate~\cite{Akarsu:2023mfb,Akarsu:2024eoo}. Taken together, these results indicate that $\Lambda_{\rm s}$CDM is a promising extension of $\Lambda$CDM in the sense of providing a single underlying mechanism---a mirror AdS-to-dS transition (or a similar phenomenon) around $z_\dagger\sim 2$---that can mitigate multiple late-time tensions simultaneously when the data favor sufficiently low $z_\dagger$.

To summarize our observational findings, we first constrained both models using the PL18 dataset alone, where the only statistically significant tension appeared in $\gamma\Lambda$CDM, with the $H_0$ tension persisting at $4.1\sigma$, while $\gamma\Lambda_{\rm s}$CDM remained free of statistically significant tensions. Including the $f\sigma_8$ data confirmed the emergence of the $\gamma$ tension in $\gamma\Lambda$CDM, with $\Omega_{\rm m0} \approx 0.31$ and $\gamma \approx 0.64$ at $3.7\sigma$ significance, consistent with Ref.~\cite{Nguyen:2023fip}, while further exacerbating the $H_0$ tension to $4.2\sigma$. 
In contrast, $\gamma\Lambda_{\rm s}$CDM, with $\Omega_{\rm m0} \approx 0.28$ and $\gamma \approx 0.61$, substantially reduces both tensions, bringing the corresponding $\gamma$ and $H_0$ discrepancies down to $2.1\sigma$ and $1.1\sigma$, respectively. When incorporating the PantheonPlus data, the preferred transition redshift shifts to higher values, weakening the impact of the mirror AdS-to-dS transition, and both the $H_0$ and $\gamma$ tensions in $\gamma\Lambda_{\rm s}$CDM rise to $2.7\sigma$ and $3.0\sigma$, respectively, while in $\gamma\Lambda$CDM the corresponding tensions remain higher, at $4.4\sigma$ and $3.4\sigma$. 
Nevertheless, incorporating the PantheonPlus\&SH0ES data keeps the preferred transition in the low-$z$ regime and once again reduces both the $H_0$ and $\gamma$ tensions below $2.0\sigma$ in $\gamma\Lambda_{\rm s}$CDM, to $0.6\sigma$ and $1.8\sigma$, respectively, while in $\gamma\Lambda$CDM, both tensions persist at the $3.3\sigma$ level. The same trend is reflected in the calibration-based $M_{\rm B}$ tension, which is reduced to $0.5\sigma$ in $\gamma\Lambda_{\rm s}$CDM, compared with $3.1\sigma$ in $\gamma\Lambda$CDM. This case, namely PL18+$f\sigma_8$+PP\&SH0ES, is particularly noteworthy because it demonstrates that $\gamma\Lambda_{\rm s}$CDM provides a substantially better fit to the data. Consequently, fixing $z_\dagger$ to its best-fit value for this dataset ($z_\dagger = 2.16$) reduces the number of free parameters to match those of $\gamma\Lambda$CDM. Despite this constraint tightening, the $H_0$, $M_{\rm B}$, and $\gamma$ tensions all remain statistically insignificant, at $0.8\sigma$, $0.5\sigma$, and $1.8\sigma$, respectively.

When incorporating BAO data from SDSS or DESI into the PL18+$f\sigma_8$+PP combination, rather than SH0ES-calibrated PantheonPlus sample,
the $H_0$ tension worsens in both models, whereas the $\gamma$ tension worsens in $\gamma\Lambda_{\rm s}$CDM but remains at a comparable level in $\gamma\Lambda$CDM. In $\gamma\Lambda_{\rm s}$CDM, the $\gamma$ tension rises to $3.0\sigma$ with SDSS and $2.8\sigma$ with DESI, while the $H_0$ tension rises to $3.9\sigma$ and $3.5\sigma$, respectively, whereas in $\gamma\Lambda$CDM, these tensions remain higher, at $3.4\sigma$ and $4.4\sigma$ for SDSS, and $3.2\sigma$ and $4.4\sigma$ for DESI. 
The worsening of $\gamma\Lambda_{\rm s}$CDM arises because BAO datasets push the transition redshift to $z_\dagger > 2.29$ for SDSS and $z_\dagger > 2.40$ for DESI, causing the mirror AdS-to-dS transition to occur too early---while matter still dominates cosmic dynamics---and thereby reducing the impact of the AdS-like CC on late-time observables. This aligns with previous findings on abrupt $\Lambda_{\rm s}$CDM, where 3D BAO datasets, like those used here, weaken the model's effectiveness, in contrast to transverse (2D) BAO data, which are less model-dependent and favor lower transition redshifts, $z_\dagger<2$---see~footnote~\ref{fnote}.

Interestingly, while $\gamma\Lambda_{\rm s}$CDM provides a slightly worse fit than $\gamma\Lambda$CDM for the SDSS-extended combination, it provides a slightly better fit for the DESI-extended combination, suggesting that different BAO datasets respond differently to model assumptions. Motivated by this, we performed additional analyses using DESI data, specifically for PL18+DESI and PL18+DESI+PP\&SH0ES. In both cases, $\gamma\Lambda_{\rm s}$CDM yields high lower bounds on the transition redshift, $z_\dagger\gtrsim2.4$, and provides a better fit than $\gamma\Lambda$CDM, although the improvement is only marginal for PL18+DESI and becomes substantial for PL18+DESI+PP\&SH0ES. The corresponding tensions in $\gamma\Lambda_{\rm s}$CDM are reduced to $\lesssim1.5\sigma$ for $\gamma$, while the $H_0$ tension remains at $3.3\sigma$ in both cases, compared to $\sim3\sigma$ for $\gamma$ and $\gtrsim3.5\sigma$ for $H_0$ in $\gamma\Lambda$CDM. These improvements do not have the same origin in the two cases: for PL18+DESI, the alleviation in the $\gamma$ tension is driven largely by the increased uncertainty on $\gamma$, whereas for PL18+DESI+PP\&SH0ES it is accompanied by a substantial improvement in fit.

Finally, we examined the special case of fixing $z_\dagger = 1.7$, motivated by previous studies on abrupt $\Lambda_{\rm s}$CDM that identified this value as a sweet spot for simultaneously resolving multiple major tensions~\cite{Akarsu:2022typ,Akarsu:2023mfb,Akarsu:2024eoo,Yadav:2024duq}. Although ${z_\dagger = 1.7}$ is not statistically favored by most of the dataset combinations considered in our analysis, it remains consistent with the lower bound ${z_\dagger > 1.58}$ found for Planck-alone and PL18+$f\sigma_8$, which led us to focus on these two cases. 
In this scenario, $\gamma\Lambda_{\rm s}$CDM, yielding $\Omega_{\rm m0} \approx 0.27$, underscores the key mechanism of the model by bringing both the $\gamma$ and $H_0$ tensions below $1\sigma$ for PL18 and, for PL18+$f\sigma_8$, reducing the $H_0$ tension to $0\sigma$ with $H_0 = 73.06 \pm 0.73\,\mathrm{km\,s^{-1}\,Mpc^{-1}}$ and the $\gamma$ tension to $1.6\sigma$ with $\gamma = 0.586 \pm 0.022$, indicating consistency with the standard $\Lambda_{\rm s}$CDM framework, while keeping $S_8 \approx 0.77$, consistent with the value inferred from KiDS-1000 in standard $\Lambda_{\rm s}$CDM analyses~\cite{Akarsu:2023mfb,Akarsu:2024eoo}.
Although this fixed-redshift benchmark yields a statistically worse fit to the data, it still provides a useful demonstration that, near the sweet spot $z_\dagger\approx1.7$ identified in standard ($\gamma=0.55$) $\Lambda_{\rm s}$CDM studies, a late-time mirror AdS-to-dS transition can alleviate the major cosmological tensions---in $H_0$, $S_8$, and the recently identified $\gamma$ tension---within these restricted dataset combinations.

Our current work is based on the simplest, idealized version of the $\Lambda_{\rm s}$CDM framework---namely, the abrupt $\Lambda_{\rm s}$CDM model---where an abrupt AdS-to-dS transition is adopted phenomenologically as a proxy for a rapid cosmic transition under the assumption that gravity is governed by GR. While this provides a clear proof of concept, a fully consistent realization of the scenario would require a smooth transition, ensuring a well-defined dynamical process rather than an instantaneous change. 
Beyond being a technical refinement, such an extension would introduce distinct observational signatures depending on the transition’s duration, rapidity, and functional form, potentially leaving detectable imprints on large-scale structure growth and clustering statistics. A natural phenomenological approach would involve modeling $\Lambda_{\rm s}(z)$ with smooth functions, such as hyperbolic tangents, in redshift or scale factor while maintaining GR. Alternatively, a more fundamental route would be to establish a concrete physical mechanism for the transition. For instance, the Ph-$\Lambda_{\rm s}$CDM model~\cite{Akarsu:2025gwi}, which realizes a smooth $\Lambda_{\rm s}$CDM scenario via a phantom field with a hyperbolic $\tanh$-type potential, provides a well-motivated physical framework within GR, where the theoretical expectation for the growth index remains $\gamma \approx 0.55$. 
Furthermore, the smooth $\Lambda_{\rm s}$CDM framework has recently been embedded into modified gravity theories, where deviations from GR can alter the theoretically expected value of $\gamma$, particularly during the transition epoch. This provides an alternative direction for extending our work by systematically examining the status of the growth index tension in different physical realizations of $\Lambda_{\rm s}$CDM. Notable examples include $\Lambda_{\rm s}$VCDM~\cite{Akarsu:2024qsi,Akarsu:2024eoo}, which embeds $\Lambda_{\rm s}$CDM into VCDM, a type-II minimally modified gravity, where $\gamma$ deviations are expected primarily during the transition, and $ f(T) $-$\Lambda_{\rm s}$CDM~\cite{Souza:2024qwd}, which implements it within teleparallel $ f(T) $ gravity. 
There also exist theoretical realizations of the abrupt $\Lambda_{\rm s}$CDM scenario. $\Lambda_{\rm s}$CDM$^+$~\cite{Anchordoqui:2023woo,Anchordoqui:2024gfa,Anchordoqui:2024dqc} provides a string-inspired realization, characterized by a distinct signature---predicting a slight excess in the total effective number of neutrino species, with $N_{\rm eff} = 3.294$. Additionally, it was demonstrated in~\cite{Alexandre:2023nmh} that in various formulations of GR, an abrupt $\Lambda_{\rm s}$ can arise naturally through a change of sign in the metric signature.

Given the encouraging phenomenological performance of the $\Lambda_{\rm s}$CDM framework in alleviating multiple cosmological tensions---most notably the long-standing $H_0$ and $S_8$ discrepancies, as demonstrated in previous studies, as well as the recently identified $\gamma$ tension, as shown in the present work---further exploration of its theoretical foundations and observational consequences is well motivated. If different realizations of $\Lambda_{\rm s}$CDM continue to exhibit a comparable ability to alleviate these tensions, a systematic analysis of the growth index $\gamma$ within these scenarios would not only assess whether they encounter a $\gamma$ tension,
but also help distinguish their underlying physical mechanisms and determine which, if any, may emerge as the most viable alternative to $\Lambda$CDM.
We stress, however, that the growth-sector constraints reported here were obtained using the standard $f\sigma_8(z)$ description of RSD measurements and should therefore be interpreted in light of the caveats discussed above. In future work, it will be important to assess the robustness of the growth constraints using a fixed-comoving-scale normalization of the matter fluctuation amplitude, such as $\sigma_{12}(z)$, alongside the standard $f\sigma_8(z)$ description.

\section*{Acknowledgments}

The authors thank Enrico Specogna for useful discussions. \"{O}.~A.\ acknowledges the support of the Turkish Academy of Sciences in the scheme of the Outstanding Young Scientist Award (T\"{U}BA-GEB\.{I}P). L.~A.~E.\ acknowledges partial financial support from the T\"{u}rkiye Bilimsel ve Teknolojik Ara\c{s}t{\i}rma Kurumu (T\"{U}B\.{I}TAK, Scientific and Technological Research Council of T\"{u}rkiye) through Grant No.\ 124N627. E.~D.~V. is supported by a Royal Society Dorothy Hodgkin Research Fellowship. J.~A.~V. acknowledges the support provided by UNAM-DGAPA-PAPIIT IN109126, IN110325, HTC project LANCAD-UNAM-DGTIC-477 and the Cátedra de Investigación Marcos Moshinsky. This article is based upon work from the COST Action CA21136 ``Addressing observational tensions in cosmology with systematics and fundamental physics'' (CosmoVerse), supported by COST (European Cooperation in Science and Technology). We acknowledge IT Services at The University of Sheffield for the provision of services for High
Performance Computing.

\section*{Data Availability}

The data that support the findings of this article are openly available~\cite{Planck:2018nkj, Said:2020epb,Beutler:2012mnras,Huterer:2016uyq,Boruah:2019icj,Turner:2022mla,Blake:2011mnras,Blake:2013nif,Howlett:2014opa,Okumura:2015lvp,Pezzotta:2016gbo,eBOSS:2020yzd, DESI:2024mwx, Scolnic:2021amr, Riess:2021jrx}.

\bibliography{struc_growth_PRD}

\end{document}